\documentclass[pre,twocolumn,superscriptaddress,aps,notitlepage,10pt]{revtex4-1}
\usepackage[
  bookmarks=true,
  colorlinks,
  linkcolor=blue,
  urlcolor=blue,
  citecolor=blue,
  plainpages=false,
  pdfpagelabels,
  final,
  breaklinks=true
]{hyperref}
\usepackage[utf8]{inputenc}
\usepackage{listings}
\usepackage{color}
\usepackage{amsmath}
\usepackage{amsfonts}
\usepackage{graphicx}
\usepackage{braket}
\usepackage{subfig}
\usepackage[normalem]{ulem}
\usepackage{cprotect}

\definecolor{fuchsia}{rgb}{1.0, 0.0, 1.0}

\begin{document}
\title{Towards a scalable discrete quantum generative adversarial neural network}
\author{Smit Chaudhary}
\affiliation{Menten AI, San Francisco, CA 94111, USA}
\affiliation{Delft University of Technology, Delft, The Netherlands}
\author{Patrick Huembeli}
\affiliation{Menten AI, San Francisco, CA 94111, USA}
\author{Ian MacCormack}
\affiliation{Menten AI, San Francisco, CA 94111, USA}
\author{Taylor L. Patti}
\affiliation{NVIDIA, Santa Clara, CA 95051, USA}
\author{Jean Kossaifi}
\affiliation{NVIDIA, Santa Clara, CA 95051, USA}
\author{Alexey Galda}
\affiliation{Menten AI, San Francisco, CA 94111, USA}
\date{\today}

\begin{abstract}
   We introduce a fully quantum generative adversarial network intended for use with binary data. The architecture incorporates several features found in other classical and quantum machine learning models, which up to this point had not been used in conjunction. In particular, we incorporate noise reuploading in the generator, auxiliary qubits in the discriminator to enhance expressivity, and a direct connection between the generator and discriminator circuits, obviating the need to access the generator's probability distribution. We show that, as separate components, the generator and discriminator perform as desired. We empirically demonstrate the expressive power of our model on both synthetic data as well as low energy states of an Ising model. Our demonstrations suggest that the model is not only capable of reproducing discrete training data, but also of potentially generalizing from it.
\end{abstract}

\maketitle

\section{Introduction}
Unsupervised and semi-supervised machine learning models have been successful in recent years at learning from and generating complex sets of data, ranging from images~\cite{xie2017aggregated, yu2022coca, ramesh2021zero}, to audio~\cite{vasquez2019melnet}, to time series data~\cite{ismail2019deep}, and more. These models typically involve training on a set of data whose essential features are distilled into a more compact form, from which new samples that share the qualitative features of the training data can be generated. Architectures for such generation include variational autoencoders, generative adversarial networks (GANs), and language models like transformers \cite{2014arXiv1406.2661G}. In this paper, we focus on GANs. Generative adversarial models such as the style GAN \cite{2018arXiv181204948K, karras2020analyzing, karras2021alias} demonstrate state of the art image generation capabilities. GANs also may be particularly well-suited as an architecture for unsupervised quantum machine learning, which we discuss in more detail below. Specifically, we introduce a quantum GAN architecture for learning from discrete datasets, and show that this architecture has properties that suggest its scalability to larger system sizes.

GANs come in a variety of forms, but all consist of two basic components: a generator (G), which attempts to generate new samples from the distribution of training data, and a discriminator (D) that attempts to distinguish the fake data from $G$ from the real training data. During training, the two components (which usually take the form of neural networks) compete in an adversarial minmax game in which convergence is reached when the discriminator can no longer distinguish between the real data and the generator's forgeries. The optimization is expressed generically in the following form:
\begin{gather}
    \min_G \max_D \lbrack  \langle \log \left(D(x) \right) \rangle_{x \sim p_{data}(x)}\\ 
     -\langle \log \left(1-D(G(z)) \right) \rangle_{z \sim p_{z}(z)}\rbrack,\nonumber
    \label{generic_loss}
\end{gather}
where $p_{data}(x)$ is the training data distribution, and $p_z(z)$ is a random distribution (often a multivariate Gaussian distribution, called the prior), which the generator learns to map to samples that resemble the data, $x$. This function is derived from the Jensen-Shannon divergence~\cite{2014arXiv1406.2661G}, and quantifies the extent to which the data distribution and the distribution of samples from the generator overlap. The first term is maximized when $D$ correctly assigns labels to all `real' and `fake' (generated) data, while the second one is minimized when $G$ produces samples that $D$ classifies as real data (the functions, $D$ and $G$ have a probabilistic interpretation here). We update $D$ to maximize the former and $G$ to minimize the latter, in an alternating, adversarial optimization. At the end of a successful training run, $D$ should misclassify samples from $G$ half the time.

Since a GAN requires sampling fake data from a probability distribution, $p(z)$, it is natural to wonder if a parameterized quantum circuit would be well-suited as a generator. After all, a wave function naturally encodes a potentially very intricate probability distribution over basis states, and the measurement process naturally corresponds to taking a sample from this distribution in a very efficient way. Indeed, sufficiently large quantum circuits can very easily encode probability distributions that cannot be efficiently sampled on a classical computer~\cite{2018arXiv181011922D}. This suggests the possibility of a more expressive \emph{quantum} generator, which could produce a more intricate array of samples than, for example, a classical neural network with a Gaussian prior distribution.

In this paper, we take advantage of this increased generative capability by combining a quantum generator with a quantum discriminator into a fully quantum GAN (QGAN), which we specifically apply to discrete data sets encoded as binary strings. We first review recent work on partially and fully-quantum GAN architectures. We then introduce the separate components of our QGAN architecture, contrasting them with previous architectures and demonstrating their capabilities as independent components. We then explain how the full QGAN is trained. We demonstrate the full QGAN by training it on the low energy states of a classical Ising model and showing that the average energy of the states sampled from the generator is lower than that of a random sample, which serves as evidence of generalization. Finally, we conclude with a discussion about the scalability of the QGAN and future work.
\section{Previous work on GANs and QGANs}
\label{review}
The recent emergence of quantum machine learning~\cite{schuld2021machine} has lead to the proposal of various hybrid architectures, including quantum generative adversarial networks (QGANs) \cite{dallaire2018quantum, zoufal2019quantum, zeng2019learning, situ2020quantum, huang2021experimental, stein2021qugan, beer2021dissipative, niu2022entangling, borras2022impact, chang2022running}. Compared to other probabilistic models, GANs are especially well suited for implementation on quantum devices because they do not require exhaustive sampling during training. Even more importantly, they do not require access to the output probability distribution of the generator. These attributes render them a promising candidate for a scalable generative quantum model. For a QGAN, we replace the classical functions $G$ and $D$ from the previous section with parameterized quantum circuits $U_G(\theta_G)$ and $U_D(\theta_D)$. 

Previous works on QGANs have explored different architectures, benchmarked on both binary and continuous data. In~\cite{zoufal2019quantum,stein2021qugan,huang2021experimental,rudolph2022generation}, the authors focus on continuous data, obtained by measuring expectation values at the output of the quantum generator. As we will later see, their architectures cannot be ported directly to our binary data. Furthermore, in~\cite{zoufal2019quantum,rudolph2022generation,borras2022impact,chang2022running} as well as to a lesser extent in~\cite{huang2021experimental}, the authors use a classical discriminator. In~\cite{beer2021dissipative}, the authors focus on learning 1-qubit quantum states.
QGANs for binary data have been studied in~\cite{zeng2019learning,situ2020quantum,borras2022impact}. In these schemes the binary samples are obtained via single-shot measurements of the generator's output. Both proposals use a classical discriminator. Since single shot samples are not as readily differentiable as expectation values, the binary generator has to be trained via the log-likelihood of its output probability distribution. This renders the method computationally intractable, as it scales exponentially in the number of qubits, requiring the estimation of $\mathcal{O}(2^N)$ output probabilities for $N$ qubits.

For continuous data, the output of the generator circuit is obtained by measuring expectation values of arbitrary $n$-qubit operators. A common choice are Pauli-$Z$ operators measured on each qubit. If such a continuous valued quantum generator is employed alongside a classical discriminator, the expectation values are used as the input to the discriminator. Since expectation values are differentiable, the training of the quantum-classical hybrid circuit can be performed via gradient backpropagation~\cite{backprop}.

In contrast, with a quantum discriminator, the generator and the discriminator form a single quantum circuit where the output state of the generator is directly fed into the discriminator. As a result, no measurements are required at the output of the generator and only a single qubit measurement is required at the output of the discriminator to obtain the label `fake' or  `real'.

In other words, using a classical discriminator with a discrete quantum generator somewhat defeats the purpose of a Quantum GAN. One of the advantages of using a GAN is that the output probabilities of the generator $p_G(x)$ do not have to be measured, which is particularly important when $p_G(x)$ corresponds to the amplitudes of a very large-dimensional wave function. For a batch of discrete samples $x \sim p_G(x)$ and a classical discriminator function $y = f_D(x)$, the average output of the discriminator with respect to a batch of samples is $\langle y \rangle = \frac{1}{M} \sum_{x \sim p_G(x)} f_D(x)$, where $M$ is the number of samples $x$. To take the gradient of $\langle y \rangle$ with respect to the parameters $\theta_G$ of $p_G$ one has to employ the log-likelihood `trick', i.e.
\begin{align}
    \nabla_{\theta_G} \langle y \rangle = \frac{1}{M} \sum_{x \sim p_G(x)} \nabla_{\theta_G}\log(p_G(x)) f_D(x).
\end{align}
Measuring the probabilities $p_G(x)$ and their derivatives would require an exponential amount of measurements. Thus, employing a classical discriminator with a quantum generator for discrete data is not a scalable approach.

To our knowledge, none of the proposed QGAN architectures for discrete data use a quantum discriminator, relying instead on measuring the sample probabilities $p_G(x)$. However, there is good reason to believe that generative quantum models are well-suited to the task of learning and generalizing from distributions of discrete data~\cite{2021arXiv210106250A, 2022arXiv220713645G, gili2022evaluating, banchi2021generalization}. This juxtaposition motivates the central question of this work: can discrete quantum GANs be constructed with a quantum generator and a quantum discriminator and trained on the discriminator labels alone? Furthermore, the aim of this work is to better understand the capabilities of QGANs to learn binary data distributions. To test the components of our QGAN architecture, we use a synthetic ``bars and stripes dataset'', which provides a simple and easily visualizable check as to whether the generator and discriminator are functioning properly. We then demonstrate the QGAN by training it on a subset of the low energy states of classical Ising Hamiltonians. Hamiltonians of this sort are particularly relevant, as they are used to encode quadratic unconstrained binary optimization (QUBO) problems of the sort solved by the quantum approximate optimization algorithm (QAOA). As such optimization problems are vital to a wide variety of fields, if the QGAN can meaningfully supplement algorithms like QAOA by providing additional low energy solutions, it would impact numerous practical applications. 
\section{Our QGAN Architecture}
We here introduce the components of our QGAN architecture that distinguish it from other proposed QGANs. Before explaining the details of our generator and discriminator, we first note that we use a Wasserstein-type loss~\cite{arjovsky2017wasserstein} to train our GAN, rather than the standard GAN loss given in Eq.~\ref{generic_loss}. Generically, the Wasserstein loss replaces non-linear activations at the end of the discriminator with linear ones:
\begin{gather}
    \min_G \max_D\quad  \langle D(x) \rangle_{x \sim p_{data}(x)}- \langle D(G(z)) \rangle_{z \sim p(z)}.
    \label{Eq:WGAN_loss}
\end{gather}
Wasserstein-type loss functions makes use of the earth movers (EM) distance metric instead of other popular probability distances used to learn distributions. This allows for much more stable training, and mitigate issues such as the vanishing gradient that are encountered with the standard GAN loss function. Furthermore, the measurement expectation values can be directly used in our loss function. In our case the EM distance is applied to the classical expectation values and, therefore, should not be confused with the quantum EM from~\cite{kiani2021quantum}.
\subsection{Training Data}
To train a generative model, we require a set of training data points $\mathcal{D} = \{ x^{(k)} \}_{k=1}^{N_{data}}$ that are sampled from the data distribution $x^{(k)} \sim p_{data}(x)$, which constitutes the ground truth that the GAN should learn.
\subsubsection{Bars and Stripes}
In this manuscript, we focus on two datasets that we want the QGAN to reproduce. The first one is the so-called `Bars and Stripes' (B\&S) dataset, which was also used in~\cite{benedetti2019generative}. It consists of binary square matrices, each with either its columns or rows set to $1$ or $0$. We focus on the $2\times 2$ B\&S dataset with 4 pixels that consists of $N_{data} = 6$ images to be learned. All other binary configurations of the $2^4$ dimensional state space are considered noise. In this simple case of B\&S, the dataset $\mathcal{D}$ contains all the training data and the data distribution $p_{data}(x) = 1/6$ is uniform. We will refer to the training data by their `image index' which are $0,3,5,10,12$ and $15$. This comes from the fact that the binary representation of e.g. $3$ is \verb|0011| which is a `bar' if it is reshaped to a 2x2 image.
\subsubsection{Ising Data}
The second dataset consists of low energy states of classical Ising-like Hamiltonians. Specifically, we consider the Hamiltonian
\begin{equation}
    H= \sum_{i,j=1}^N J_{ij} Z_i Z_j + \sum_{i=1}^N h_iZ_i.
    \label{isingham}
\end{equation}
This dataset is inspired by the fact that many real-world optimization problems can be formulated as Ising Hamiltonians~\cite{lucas2014ising}, with some of the more prominent applications including protein design and drug discovery~\cite{mulligan2020designing}, the travelling salesman~\cite{jain2021solving} and portfolio optimization in finance~\cite{herman2022survey}.

We here consider the simplest case of Eq.~\ref{isingham}: constant, nearest-neighbor interactions, zero magnetic field, and open boundary conditions, producing the Hamiltonian
\begin{equation}
    H_{NN}= \sum_{i=1}^{N-1}  Z_i Z_{i+1}.
    \label{simple_ising}
\end{equation}
The dataset $\mathcal{D}$ consists of uniformly drawn low energy states of $H_{NN}$.

Ultimately, the QGAN should not merely learn the training data, it should also generalize from the learned data. For example if the QGAN is trained on a subset of the low energy states of the above Hamiltonian, it should learn to generate low energy states that are not in the training dataset. Successful generalization is applicable to the more practical task of enhancing the solutions to QUBO-like problems, which take the form of the above Hamiltonian. Such diversity of solutions is often considered as crucial as finding the ground state of the Hamiltonian, particularly in drug discovery \cite{dean1999molecular, gorse_molecular_1999, galloway2010diversity}.
\subsection{The quantum discriminator}
\label{sec:quantum_discriminator}
\begin{figure}%
    \centering
    \subfloat{{\includegraphics[width=0.95\linewidth]{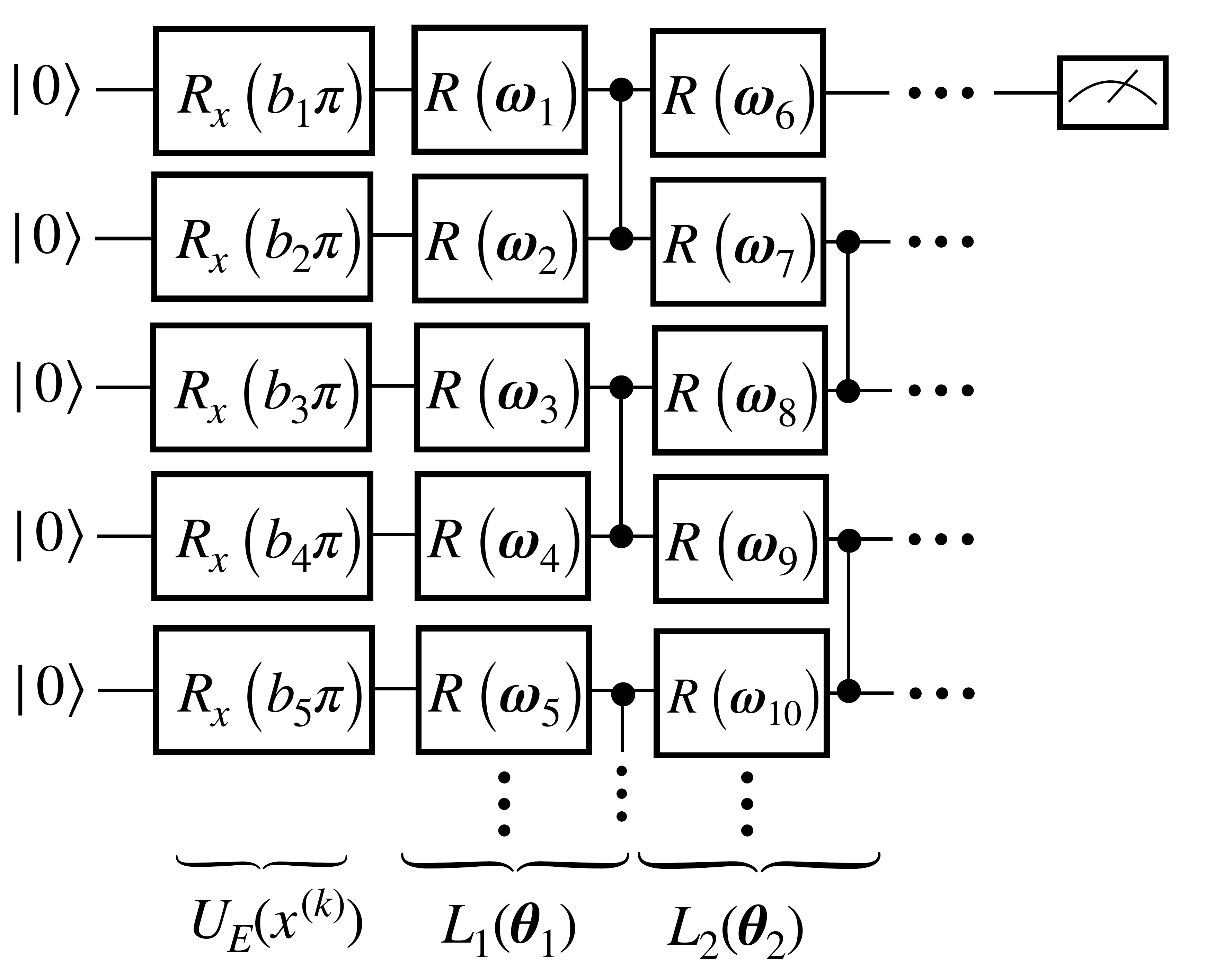} }}%
    \caption{The quantum circuit architecture of the discriminator. It consists of a layer-wise structure where the first layer is an encoding layer. I.e. it prepares the state $\ket{x^{(k)}}$ from a binary string $x^{(k)}$. The encoding is followed by Layers of parameterized single qubit rotations and CZ gates. The label of the discriminator is obtained by measuring the 1st qubit of the circuit.}%
    \label{fig:Circuit}%
\end{figure}
\begin{figure}%
    \centering
     \subfloat{{\includegraphics[width=0.95\linewidth]{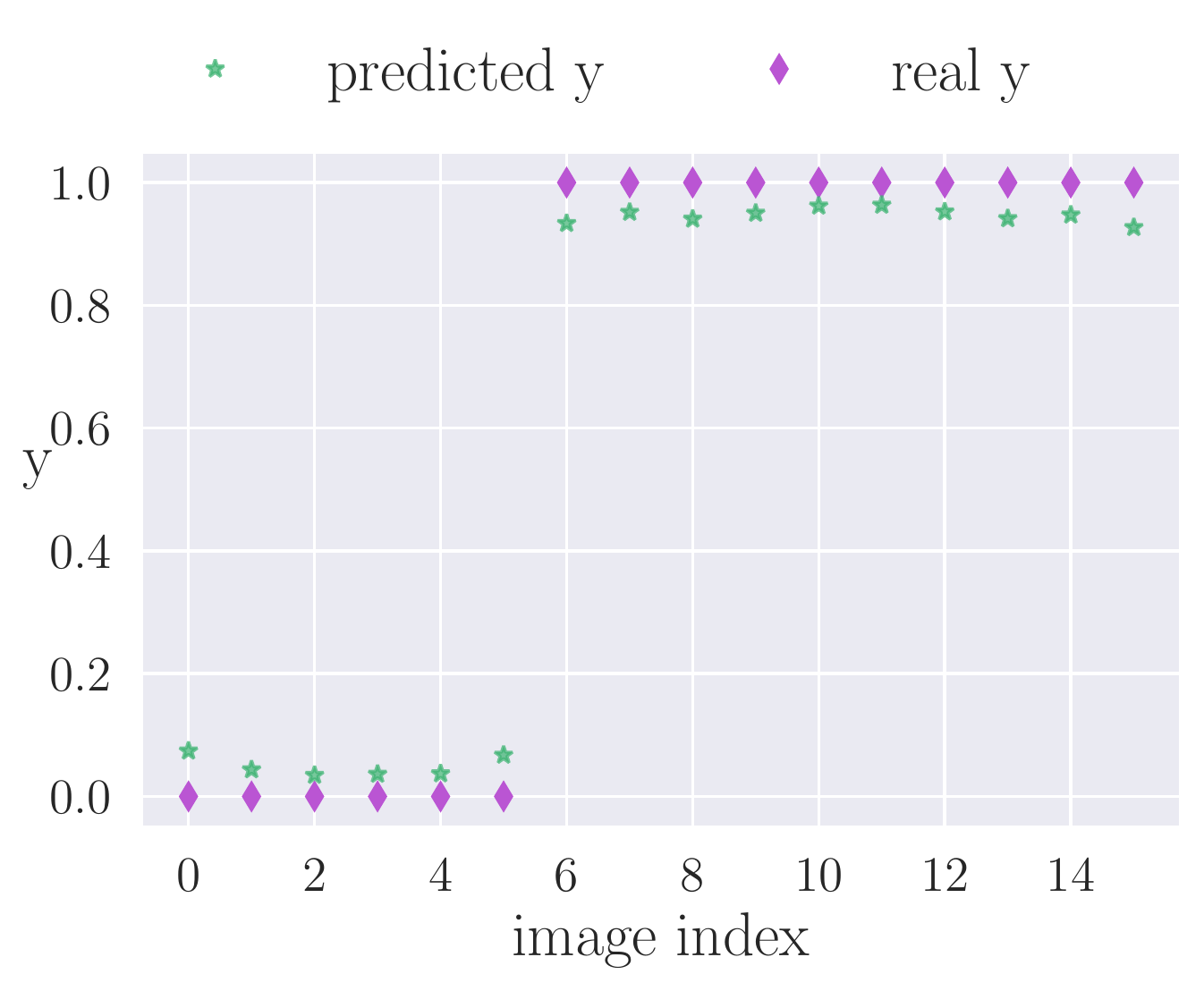} }}%
    \caption{\textbf{Discriminator on B\&S data:} The discriminator predictions $y_{pred}$ after training on the B\&S dataset with their real labels. The x-axis represents the indices of the different images of the B\&S dataset. In this figure, the `image indices' are ordered for better readability. $0$ to $5$ represent the training data. The rest of the indices represent states that are not part of B\&S.}%
    \label{fig:Discriminator_BandS}%
\end{figure}
For the training of a QGAN it is crucial to obtain a good gradient signal from the discriminator.
Therefore, we first verify that the quantum discriminator is able to distinguish the training data from any outside data.

To construct a supervised classification task for the discriminator, we take the set of binary strings $\{x^{(k)}\}$ with $x^{(k)} = (b_1, \dots, b_N)$ and $b_i \in \{0,1\}$ and add labels $y^{(k)} \in \{0,1\}$. These labels represent if a data point $x^{(k)}$ is part of the data distribution $p_{data}(x)$ or not. The aim of this task is to see if the discriminator is able to predict the correct label for a given binary input after the training.

Fig.~\ref{fig:Circuit} shows the parameterized quantum circuit architecture that we use for the discriminator.
To encode the classical data into a quantum state we use basis encoding~\cite{schuld2018supervised} which consists of layer of single qubit rotations which we denote $U_E(x) = \otimes_i^N R_x(b_i \pi)$. This layer maps $x^{(k)} \rightarrow \ket{x^{(k)}}$. The trainable, parameterized part of the circuit consists of a layerwise structure of single qubit rotations $R(\boldsymbol{\omega}_i) = R_z(\gamma_i) R_y(\beta_i) R_z(\alpha_i)$ applied to each qubit, followed by controlled $Z$ gates. 
The single qubit rotations are parameterized by the rotation angles $\alpha,~\beta$ and $\gamma$. The trainable part of the discriminator $U_D(\theta_D)$ consists of $l$ layers $L(\boldsymbol{\theta}_i)$, where $l$ is referred to as the depth of the circuit.
The variable $\theta_D$ denotes all the variational rotational angles of the discriminator circuit that are optimized over the course of training.

Finally, to obtain a label prediction $y_{pred}^{(k)}$ from the quantum discriminator, we prepare the state $U_D(\theta_D) \ket{x^{(k)}}$ and estimate the expectation value of an observable $\braket{\mathcal{O}}$, interpreting this value as the predicted label. Note that this method of retrieving the discriminator label is especially suitable for quantum circuits, as the expectation value of a single operator can be estimated efficiently, with a relatively small number of shots.

For the task at hand, i.e. distinguishing between training data and `noise', we require a simple binary discriminator. Therefore, it is sufficient
to measure a  single qubit in an arbitrary basis. In this work, we measure the first qubit of the discriminator in the Pauli $Z$ direction.
We then renormalize the measurement outcome as
\begin{align}
    y_{pred} = \frac{1}{2}\left( \braket{Z_1}  + 1\right),
\end{align}
such that the label can  later be interpreted as the probability of the input sample $x^{(k)}$ being real (i.e. that it is part of the training dataset) during training.

In the supervised setting, the discriminator circuit is trained to minimize the loss function $\frac{1}{M}\sum_k (y_{pred}^{(k)} - y^{(k)})^2$, where $M$ is the number of input label tuples $(x^{(k)}, y^{(k)})$.
We optimize the parameters of the quantum circuit by minimizing the loss function via gradient descent.

Fig.~\ref{fig:Discriminator_BandS} displays the performance of the discriminator on the B\&S data. It identifies the the B\&S data points (labeled $y=0$) and the noise data points (labeled $y=1$) with high accuracy. 

\begin{figure}%
    \centering
    \subfloat{{\includegraphics[width=0.44\linewidth]{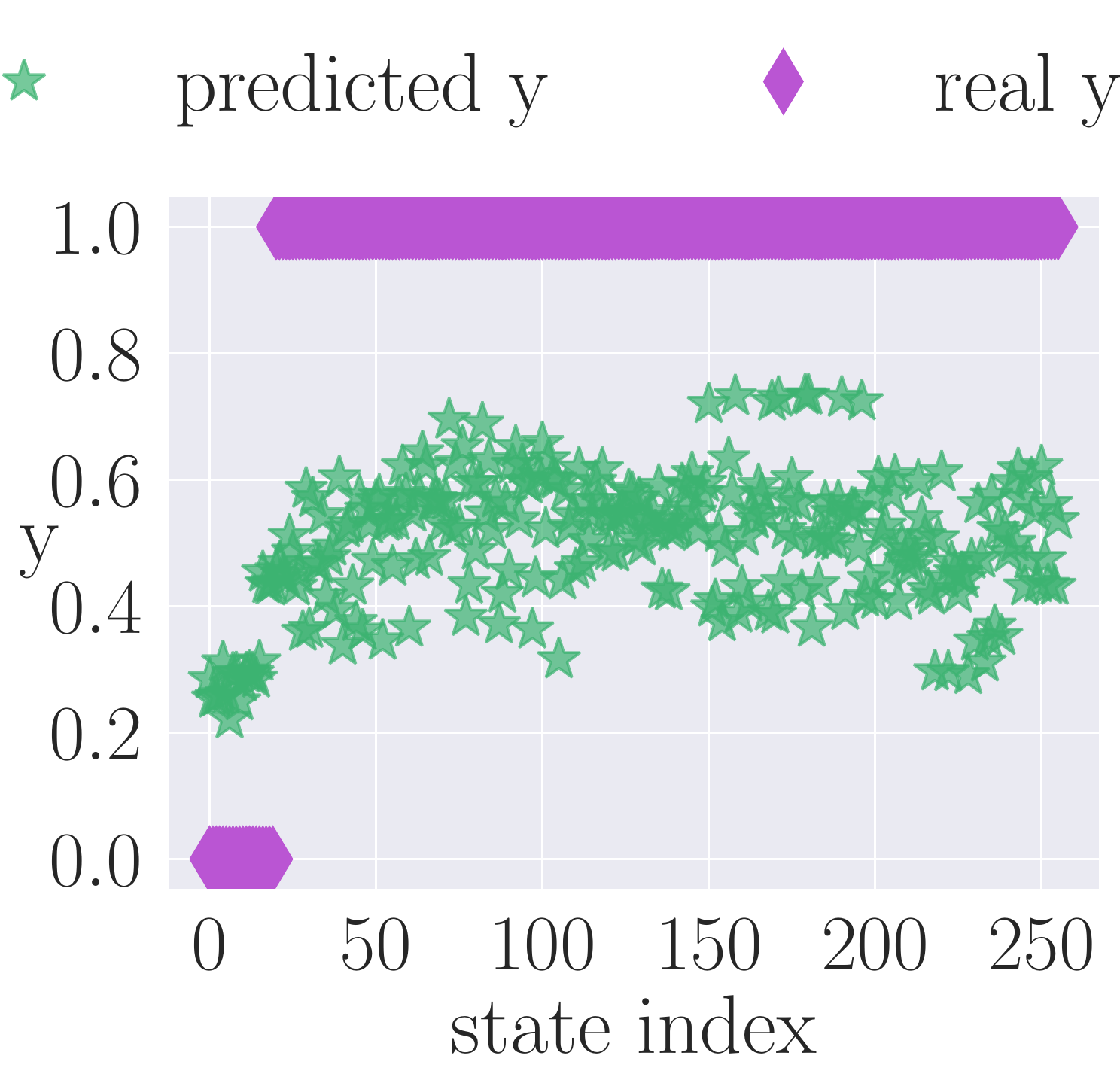} }}%
    \qquad
    \subfloat{{\includegraphics[width=0.44\linewidth]{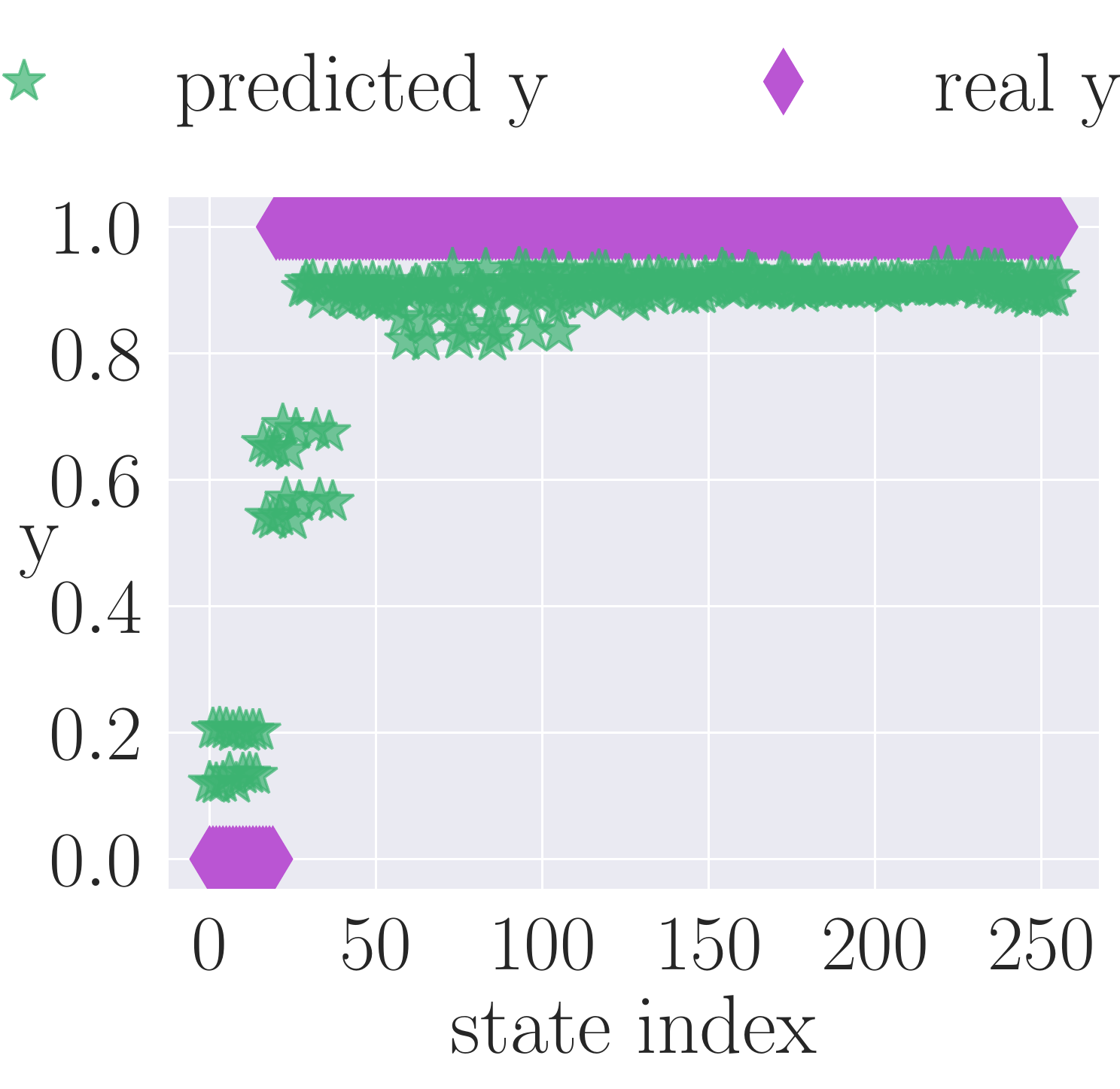} }}%
    \caption{\textbf{Discriminator on imbalanced data:} Discriminator predictions vs. real labels for the Ising data on $N=8$ spins for imbalanced datasets that contains all $2^N = 256$ states without auxiliary qubits (left) and with $N_A =4$ auxiliary qubits (right). The x-axis represents the index of the states ordered by energy.}%
    \label{fig:Discriminator_Ising_Imbalanced}%
\end{figure}

Fig.~\ref{fig:Discriminator_Ising_Imbalanced} shows the predictions of the discriminator for the Ising dataset. In a realistic scenario for the training of a QGAN one would provide a small set of low energy states as a training dataset and the discriminator needs to distinguish them from arbitrary noise samples. I.e. we benchmark the discriminator on a strongly imbalanced dataset. The dataset contains all $2^N$ energies eigenvalues of a classical Ising model with $N=8$. We split the dataset into $20$ low energy states ($y=0$) and the rest high energy states ($y=1$). The training without any auxiliary qubits (Fig.~\ref{fig:Discriminator_Ising_Imbalanced} (left panel)) converges to a solution, where the predicted labels almost can't be distinguished. If we add $N_A = 4$ auxiliary qubits the discriminator is able to label the data with higher accuracy (right panel).

This highlights an important implication of how the input data is partitioned into low and high energy states.
Namely, note that a projective measurement necessarily divides the entire Hilbert space into 2 equal parts.
For our $N=8$ spin Ising system, the discriminator necessarily divides the $256$ states into $128$ low energy states and $128$ high energy states.
This comes from the fact that for the $2^N$ states, measuring $Z$ on qubit 1 would give +1 and -1 an equal number of times.
%However, this is to do with the size of the discriminator circuit.
The measurement divides the $N$ qubit system into two equal parts, which further means that the $N$ spin Ising data is also divided accordingly, because of the unitary mapping of the binary Ising states to discriminator states.
However, it is completely reasonable to want an unequal division of the Ising states, so that the discriminator can distinguish, for example, the lowest $k$ of states from the remaining $2^N - k$.
To achieve this, we add auxiliary qubits to the discriminator circuits. These auxiliaries are initialized as $\ket{0}$, rather than according to the input data. Otherwise, the circuit architecture in Fig.~\ref{fig:Circuit} remains the same, save now with $N + N_A$ qubits, where $N_A$ is the number of auxiliary qubits.

In Appendix~\ref{Appendix:Discrimnator} we show that the discriminator is able to achieve much higher accuracy without auxiliary qubits if the dataset is balanced or also if it discriminates a set of states that is much smaller than the whole Hilbert space.
\subsection{The generator}
Since classical neural networks are fully deterministic, a classical GAN  requires a  noise vector $z$ as an input to obtain different samples $x = G(z)$. As quantum circuits are by design probabilistic and the output quantum state of the generator $\vert \psi_G \rangle  = \sum c_i \vert x_i \rangle$ can be designed such that $p(x) = \braket{x \vert \psi_G}$ is exactly $p_{data}(x)$. One could therefore conclude that noise inputs to the quantum generator are unnecessary. However, as we illustrate here, noise is also crucial for the training of binary generators, and that intrinsic quantum randomness alone is insufficient to successfully train the quantum generator. We emphasize that, while in theory the generator itself could be designed to represent the entire data probability distribution $p_{data}(x)$, in practice we do not observe the adversarial training converging to the correct distribution without a noise input.
To better understand the influence of noise on training and prescribe the generator's quantum circuit architecture, we start with two simple toy models. 

\begin{figure}%
    \centering
    \subfloat {{\includegraphics[width=0.44\linewidth]{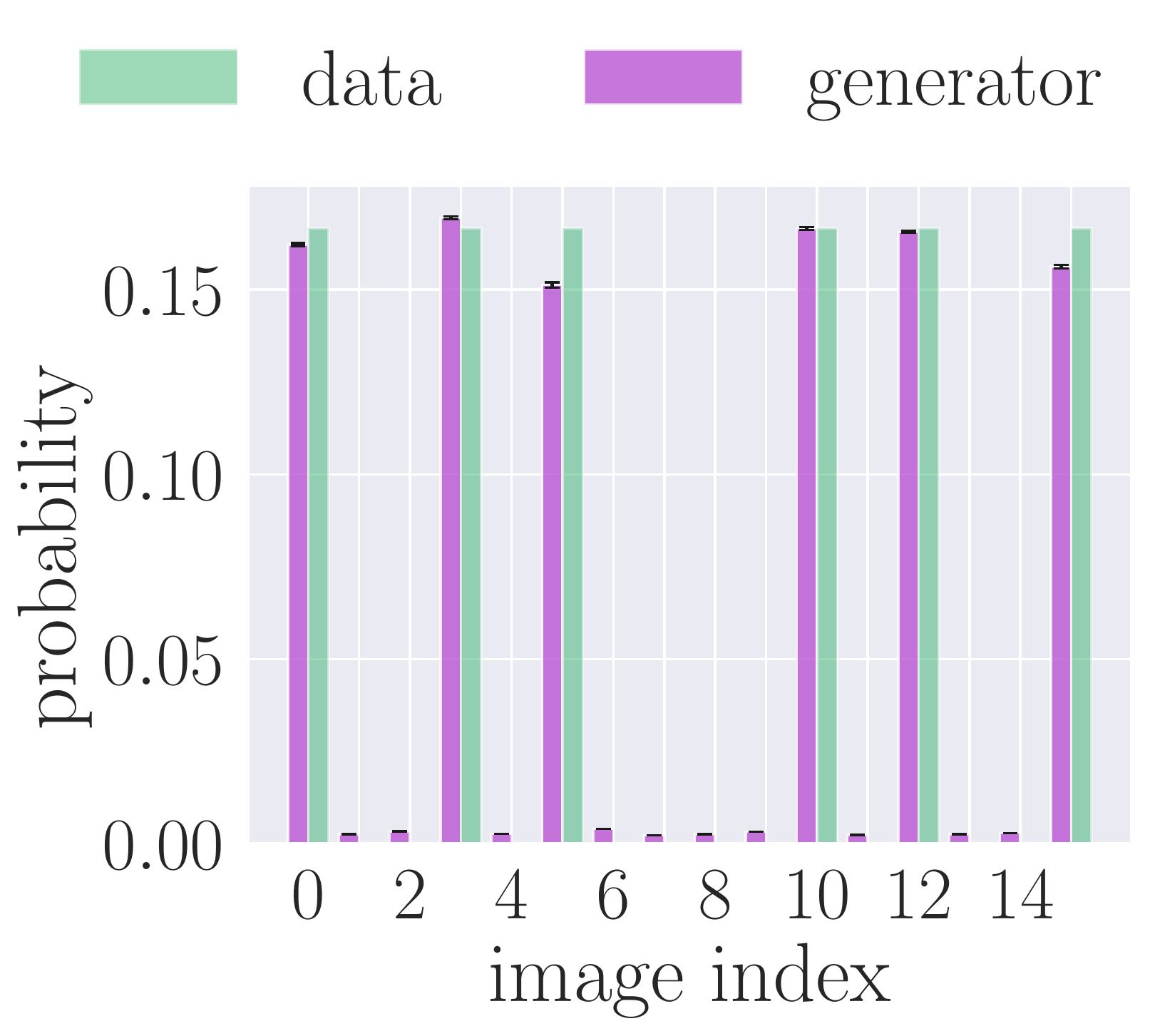} }}%
    \qquad
    \subfloat{{\includegraphics[width=0.44\linewidth]{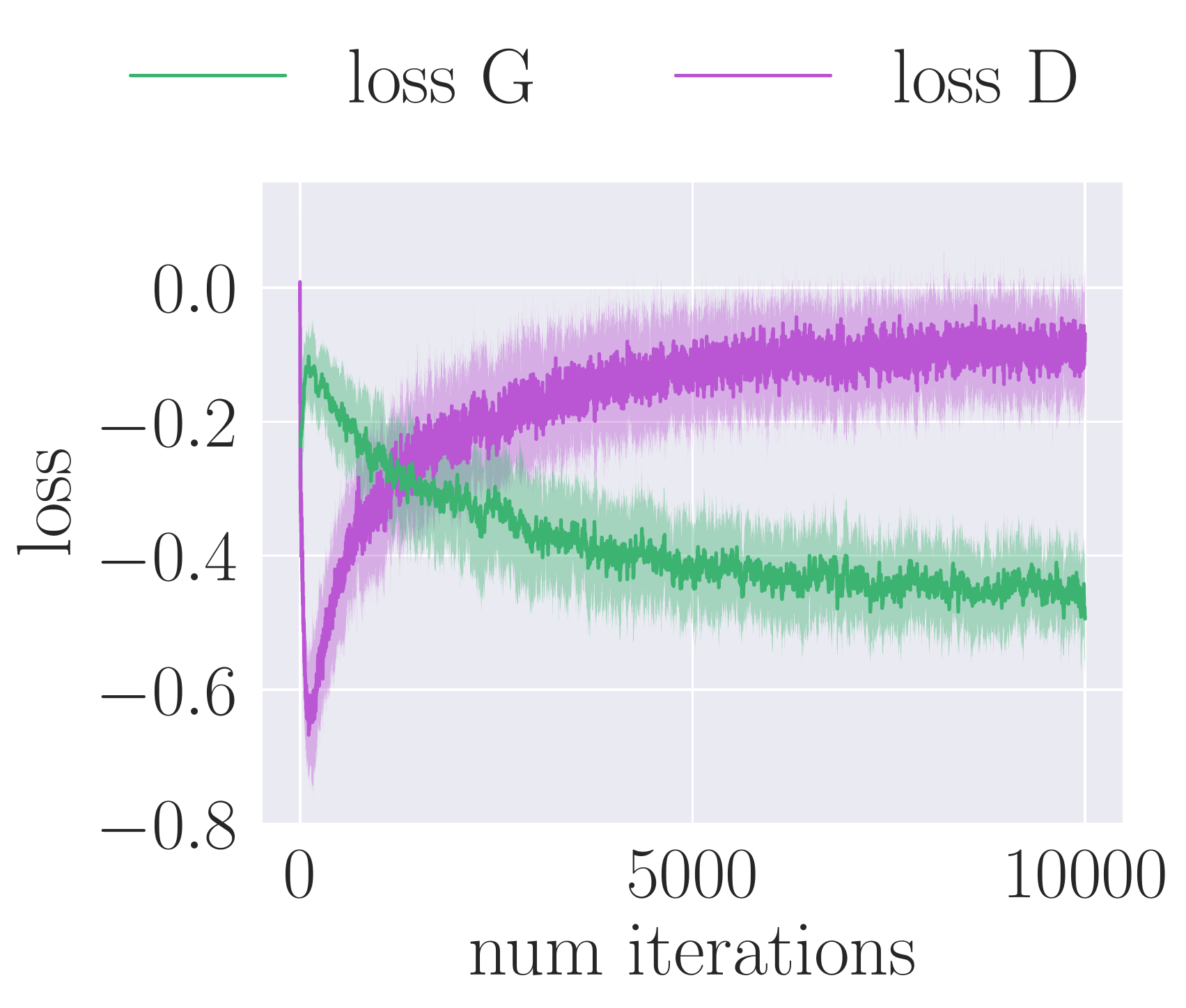} }}%
    \cprotect\caption{\textbf{Quantum generator toy model with noise:} Output distribution (left) and loss during training (right) of the generator quantum state $\vert \psi_G(z) \rangle$ with a classical noise source $z$ trained together with the discriminator circuit $U_D(\theta_D)$. The training converges as expected for a GAN and the generator learns the data. Both figures are averaged over $20$ different random initializations. The error bars in the histogram show the variance of the probabilities. The variance of the probabilities is small and barely visible which means that the $20$ training runs converge to very similar output distributions $p_G(x)$. The shaded region in the right panel shows the standard deviation.}%
    \label{fig:Toy_Model_with_noise}%
\end{figure}

First we show how the training converges if the generator state depends on a classical noise input $\vert \psi_G \rangle \rightarrow \vert \psi_G (z) \rangle$. As a simple model of such a noise conditioned state we choose a fully connected neural network (for more details see Section~\ref{Sec:Methods}) that has a noise vector $z$ as an input and a $2^N$ dimensional vector $\vert \psi_G(z) \rangle$ as an output. That is, we interpret the output of the neural network $(c_1(z), \dots c_{2^N}(z))$ as the amplitudes of the generator state $\vert \psi_G \rangle = \sum c_i(z) \vert x_i \rangle$. This simple model allows us to have a highly non-linear connection between the noise and the resulting quantum state and as a result the training losses converge to the correct values. Fig.~\ref{fig:Toy_Model_with_noise} shows the resulting loss and the output distribution of the generator $p_G(x)$ of the QGAN with a noise dependent generator state. The loss curves and the histograms are averaged over 20 different random initializations. The error bars in the histograms show the variance.

For the optimization without noise we represent the output state of the generator $\vert \psi_G \rangle = \sum c_i \vert x_i \rangle$ directly with $2^N$ trainable parameters $c_i$. In principle, this should allow us to train a generator capable of expressing the data distribution $p_{data}(x)$ and we can see if such a `perfect' classical state ansatz can be trained with the quantum discriminator $U_D(\theta_D)$ from last section. We assume $c_i \in \mathbb{R}$ without loss of generality. 

To train the QGAN we perform the optimization described in Eq.~\ref{Eq:WGAN_loss}. To estimate the expected label of the real data $\braket{D(x)}_{x \sim p_{data}}$, we prepare the state $\ket{\psi_D^{(k)}} = U_D(\theta_D) \ket{x^{(k)}}$ and take the average over the measured expectation values $\frac{1}{\vert \mathcal{B} \vert} \sum_k \bra{\psi_D^{(k)}} Z \ket{\psi_D^{(k)}}$, where $\mathcal{B}$ is random batch drawn from the training data $\mathcal{D}$. The expected label of the fake data $\langle D(G(z)) \rangle_{z \sim p(z)}$ can be estimated by preparing the state $\ket{\psi_D^{fake}} = U_D(\theta_D) \ket{\psi_G}$.

\begin{figure}%
    \centering
    \subfloat{{\includegraphics[width=0.44\linewidth]{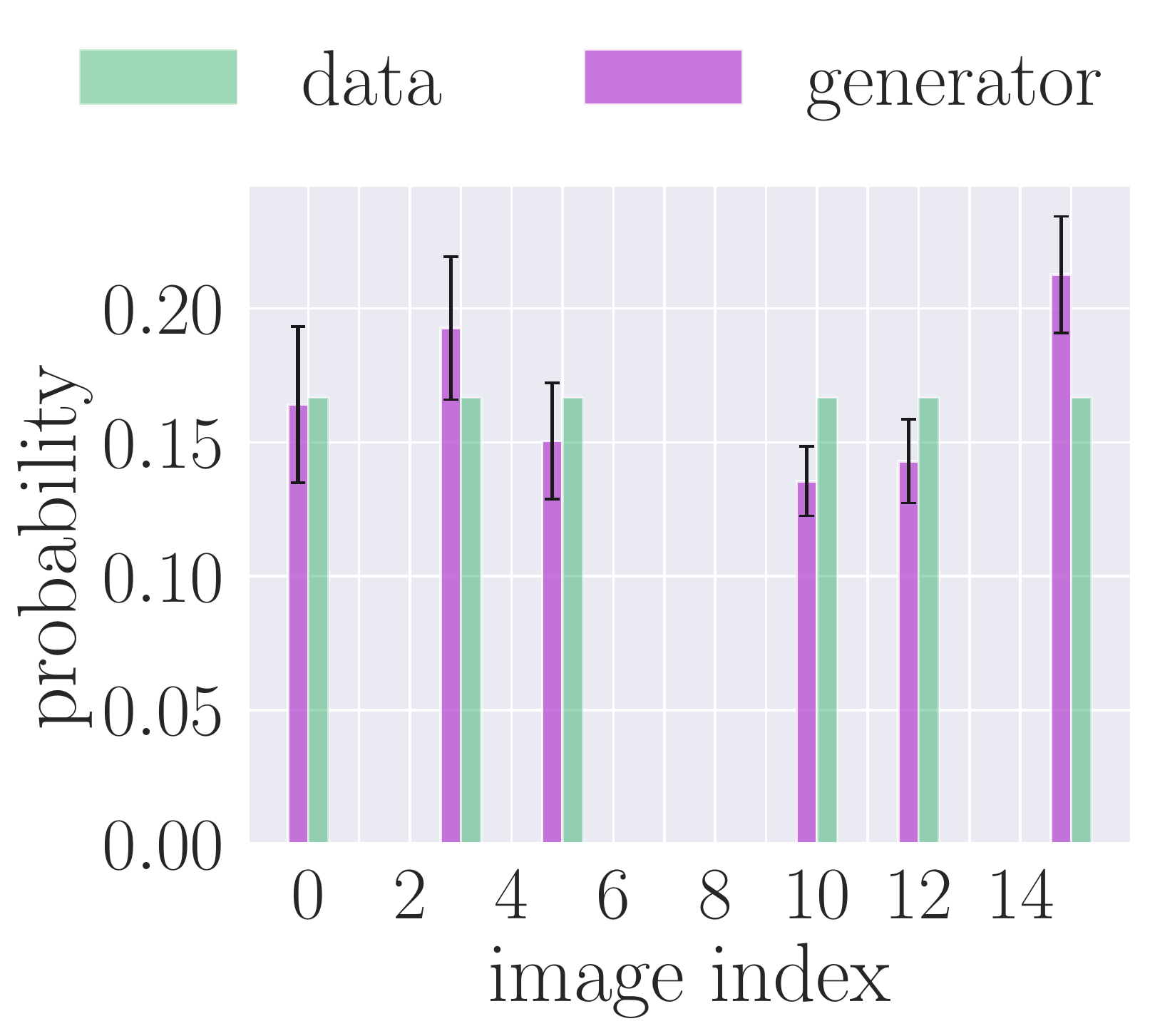} }}%
    \qquad
    \subfloat{{\includegraphics[width=0.44\linewidth]{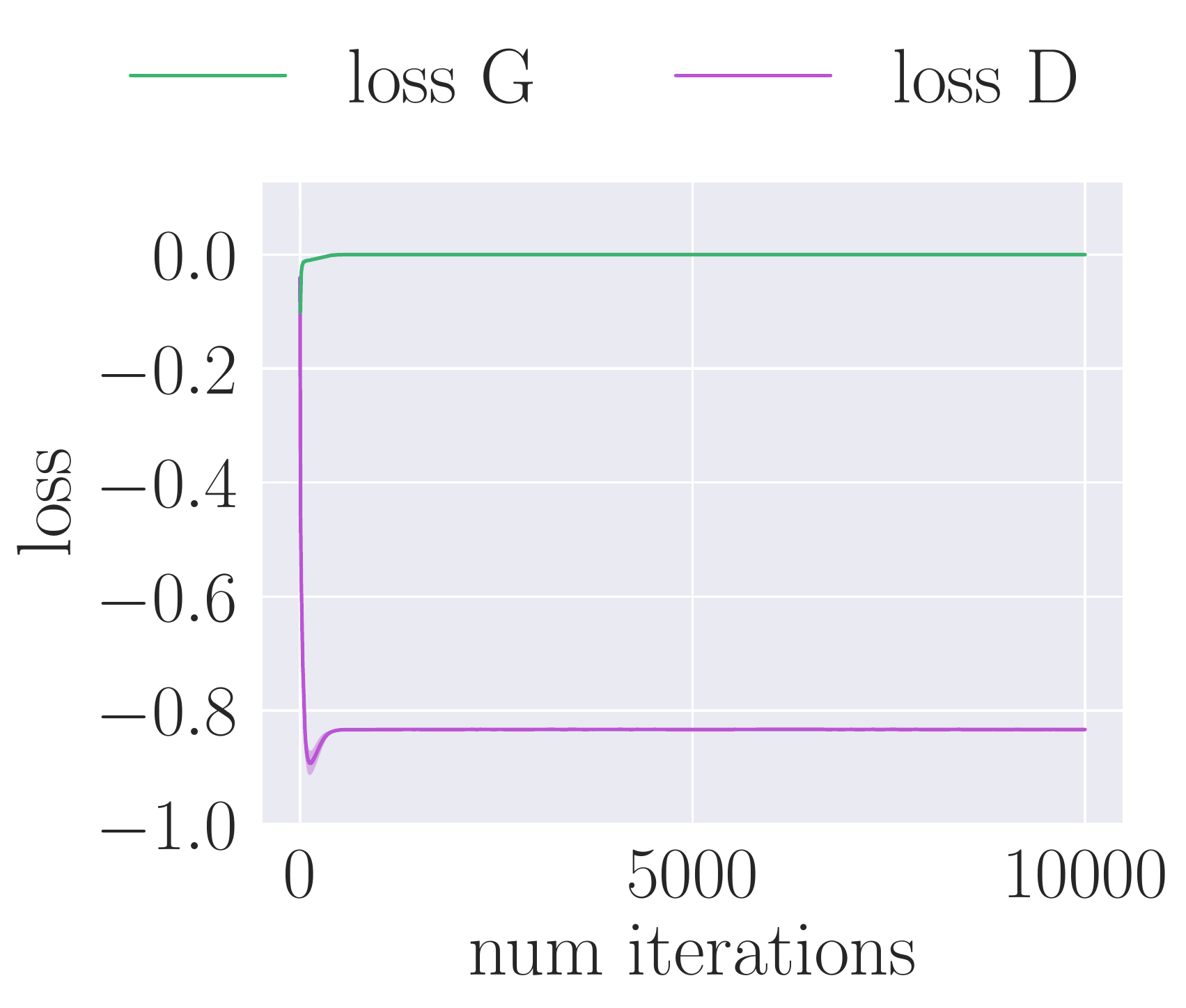} }}%
    \caption{\textbf{Quantum generator toy model without noise:} Output distribution (left) and loss during training (right) of the generator quantum state $\vert \psi_G \rangle = \sum c_i \vert x_i \rangle$ without classical noise input trained together with the discriminator circuit $U_D(\theta_D)$. The loss gets stuck very fast and the generator does not learn the training data which can be seen in the big variance of the probabilities. They manifest that the training suffers from mode collapse, where some some training data is not learned at all. An example of a single run can be found in Fig.\ref{fig:no_noise_single} in the Appendix. Both figures are averaged over 20 different random initializations. The error bars in the histogram show the variance of the probabilities. The standard deviation in the right panel is barely visible.}%
    \label{fig:Toy_Model_no_noise}%
\end{figure}

Fig. \ref{fig:Toy_Model_no_noise} shows the loss during the training and the output distribution of the generator after training $p_G(x)$.
The loss of the QGAN gets stuck in a local minima after a short amount of training. For the Wasserstein loss, the generator loss is expected to go to $loss_G = \langle D(G(z)) \rangle_{z \sim p(z)} = -0.5$, indicating that the discriminator is unable to distinguish between the fake and real data. Likewise, the discriminator loss is expected to converge to $loss_D = \braket{D(x)}_{x \sim p_{data}} - \langle D(G(z)) \rangle_{z \sim p(z)} = 0$, as the correct and incorrect classification of data should occur with equal frequency.
Even though the loss gets stuck in local minima, the distributions $p_G(x)$ and $p_{data}(x)$ are similar, but the generator tends to omit certain training samples completely and assigns too high probabilities for other training examples. This phenomenon is also called a mode collapse~\cite{2014arXiv1406.2661G}.

These toy models show that noise is a crucial ingredient to build a discrete quantum generator and that in theory it is possible to create a state $\vert \psi_G (z) \rangle$ that encodes all the training data. Since representing the generator state as a $2^N$ dimensional vector is not scalable, the toy model will be replaced with a parameterized quantum circuit. Therefore, we are looking for an appropriate circuit ansatz that maps $f: z \rightarrow \vert \psi_G (z) \rangle$. In the next section we discuss how such a quantum circuit could look like.
\subsection{Complete quantum GAN}
The output state of the quantum generator is given by circuit $\vert \psi_G(z) \rangle = U_G(\theta_G, z) \vert 0 \rangle^{\otimes N}$, where $\theta_G$ are the trainable parameters of the generator and $z$ is a noise vector coming from a classical source of noise, e.g. a uniform distribution $p_z$. The output state of the discriminator is given by a unitary $U_D(\theta_D)$ acting on the output state of the generator $\vert \psi_D(z) \rangle = U_D(\theta_D) \vert \psi_G(z) \rangle$. 
The label for a `fake' input is given by
\begin{align}
y_{\text{fake}}(z) = \frac{1}{2} \left( \langle \psi_D(z) \vert  Z  \vert \psi_D(z) \rangle + 1 \right).
\end{align}
The label for the `real' data can be obtain the same way as described in section~\ref{sec:quantum_discriminator}
\begin{align}
   y_{\text{real}}(x^{(k)}) = \frac{1}{2} \left( \langle x^{(k)} \vert U_D(\theta_D)^{\dag} Z~ U_D(\theta_D) \vert x^{(k)} \rangle + 1 \right). 
\end{align}
To evaluate a batch of classical data $\mathcal{B} = \{ x^{(k)}\}_{k=1}^M$ we take the average over single predictions $\frac{1}{M} \sum_k  y_{\text{real}}(x^{(k)})$. Another possibility would be to prepare a superposition state $\vert \psi_{\text{data}} \rangle = \frac{1}{\sqrt{M}} \sum_{k=1}^M \vert x^{(k)} \rangle$ and use this as an input state for the discriminator. In this work we focus on the former and leave the latter for further future investigation.

Previous research has shown that the so called data-reuploading circuits are universal function approximators, at least for classification tasks \cite{perez2020data, schuld2021effect}. Originally this architecture was introduced as a quantum analog to a classical neural network~\cite{perez2020data}. Therefore, it is a promising candidate to replace the classical neural network from the previous section to adopt the non-linear function that maps noise to the output state of the generator.

Data reuploading refers to a quantum circuit structure, where parameterized unitaries $U(\theta_i)$ are alternated with a data dependent unitary $U(x)$. In a layerwise structure the final circuit $U(\theta, x)$ consists of an alternating pattern of $U(\theta_i)$ and $U(x)$. In our case we do not upload data to the circuit but random noise $z$ which we refer to as noise reuploading. Therefore the generator circuit will have the form $U_G(\theta_G, z) = \prod_i U(\theta_i) U(z)$   
\begin{figure}%
    \centering
    \subfloat{{\includegraphics[width=0.44\linewidth]{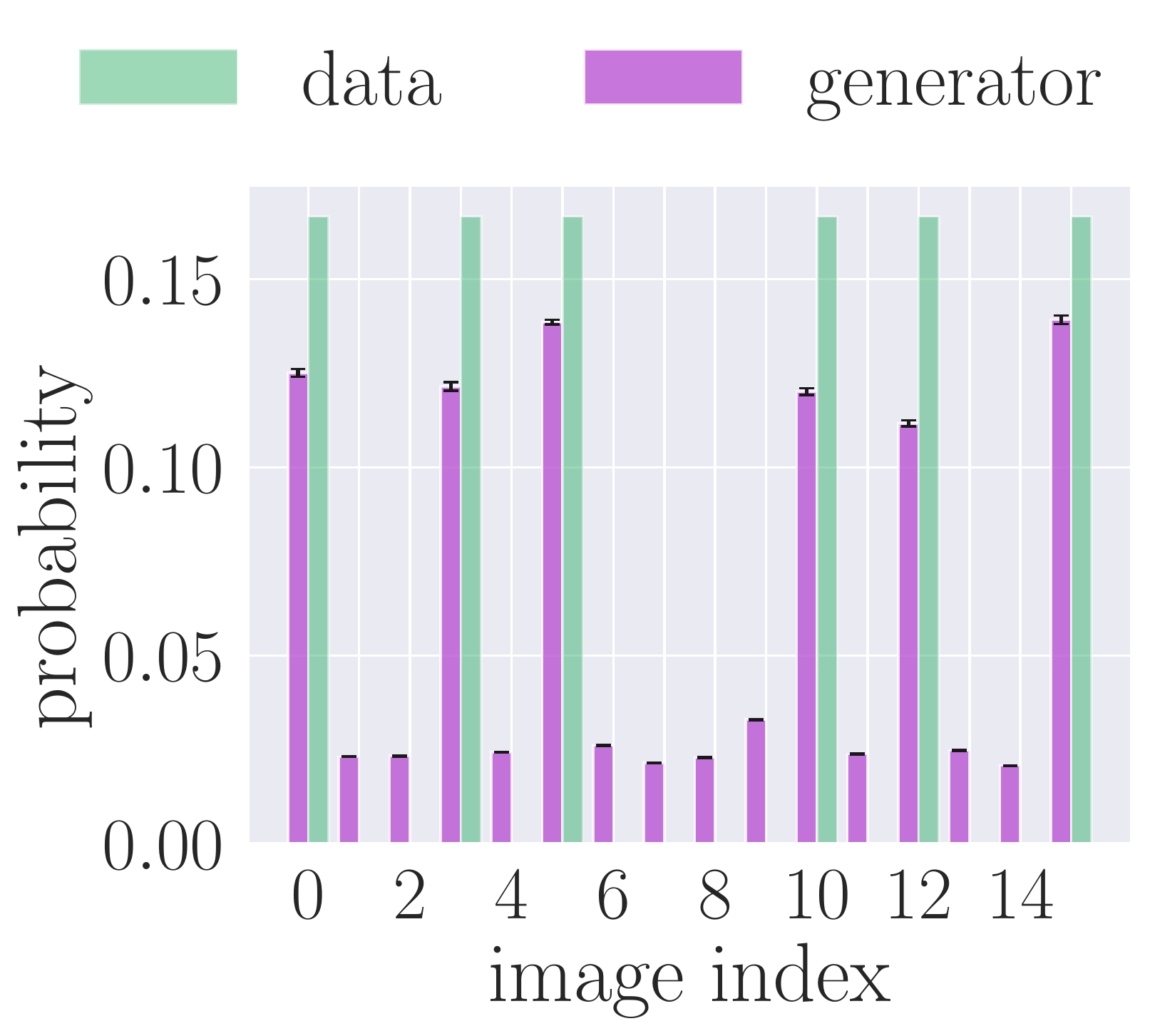} }}%
    \qquad
    \subfloat{{\includegraphics[width=0.44\linewidth]{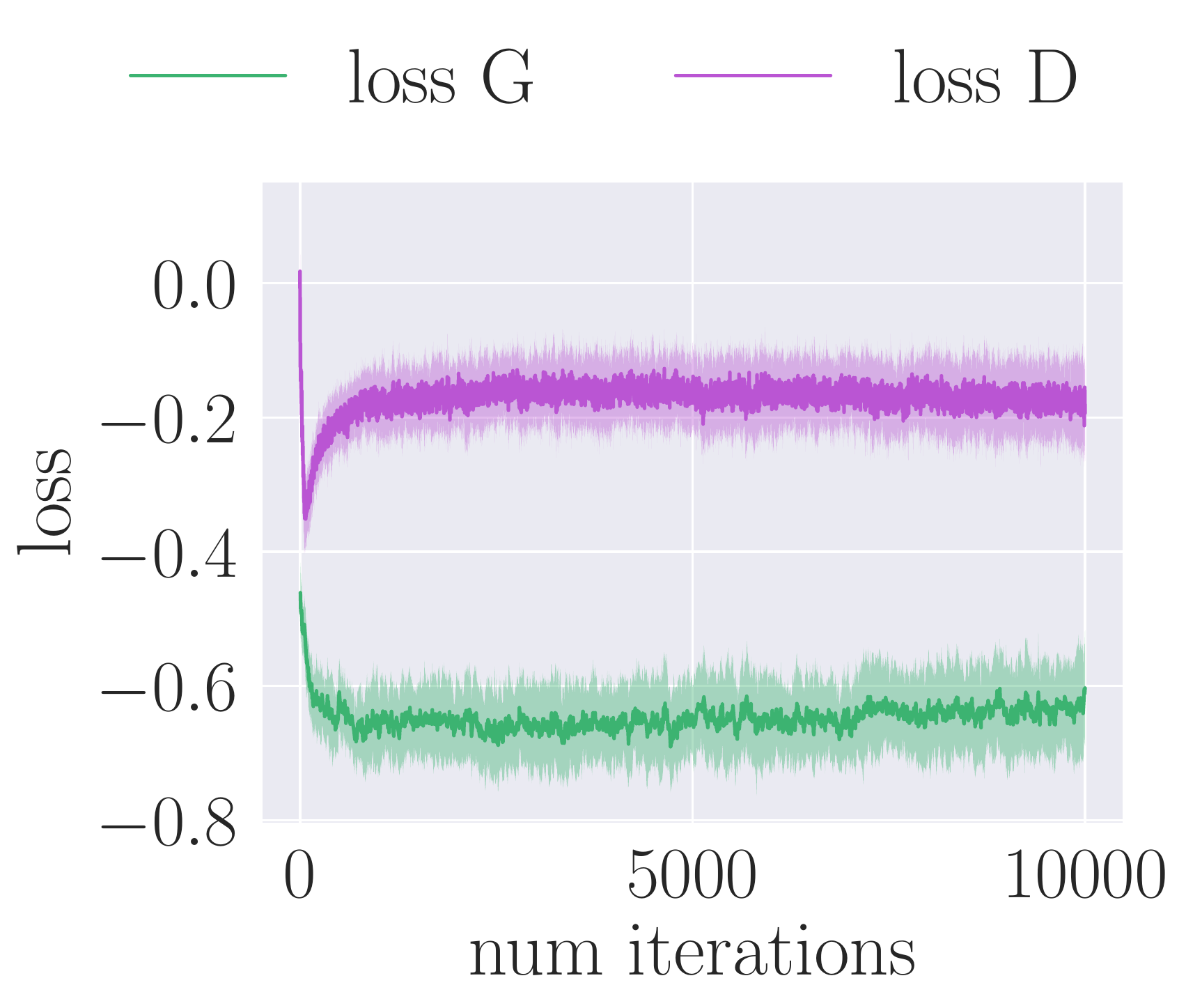} }}%
    \caption{\textbf{Quantum generator with noise reuploading:} Output distribution (left) and loss during training (right) of the generator quantum circuit with data reuploading $U_G(\theta_G, z)$ trained together with the discriminator circuit $U_D(\theta_D)$. The training converges as expected for a GAN and the generator learns the data. Both figures are averaged over 20 different random initializations. The error bars in the histogram show the variance of the probabilities. The shaded region in the right panel shows the standard deviation.}%
    \label{fig:Data_reuploading_with_noise}%
\end{figure}
Fig.~\ref{fig:Data_reuploading_with_noise} shows the training loss and the resulting output distribution of the generator for the data-reuploading circuit architecture. 

Another straight-forward way to add a source of classical noise to a parameterized part of the generator quantum circuit $U_G(\theta_G)$ is to start from an initial state $\vert z \rangle = \otimes_i^N R_{\alpha} (z_i) \vert 0 \rangle ^{\otimes N}$, with $R_{\alpha}$ an arbitrary single qubit rotation. The output state of the generator is $\vert \psi (z) \rangle = U_G(\theta_G) \vert z \rangle$. In the appendix~\ref{sec:Appendix_linear_noise} we show the training of such a circuit and that applying a unitary on the noise vector shows worse performance than the reuploading architecture. A likely conjecture is that the linear transformation that is applied to the noise vector $\vert z \rangle$ via the unitary $U_G$ does not provide a sufficiently non-linear correlation between noise and the output state $\vert \psi_G(z) \rangle$.
\section{Training QGAN on Ising Data}
In this section, we discuss training the fully quantum GAN on a low-energy subset of states from the  Ising model Hamiltonian of Eq.~\ref{simple_ising}. A central motivation for our use of QGANs in this context is their potential to generalize, i.e., to generate low energy states that are not in the training dataset. As discussed above, such generalization would have practical implications for enhancing the solutions to QUBO-like problems, due to their similarity with the above Hamiltonian.

\begin{figure}
    \centering
    \includegraphics[width=0.9\linewidth]{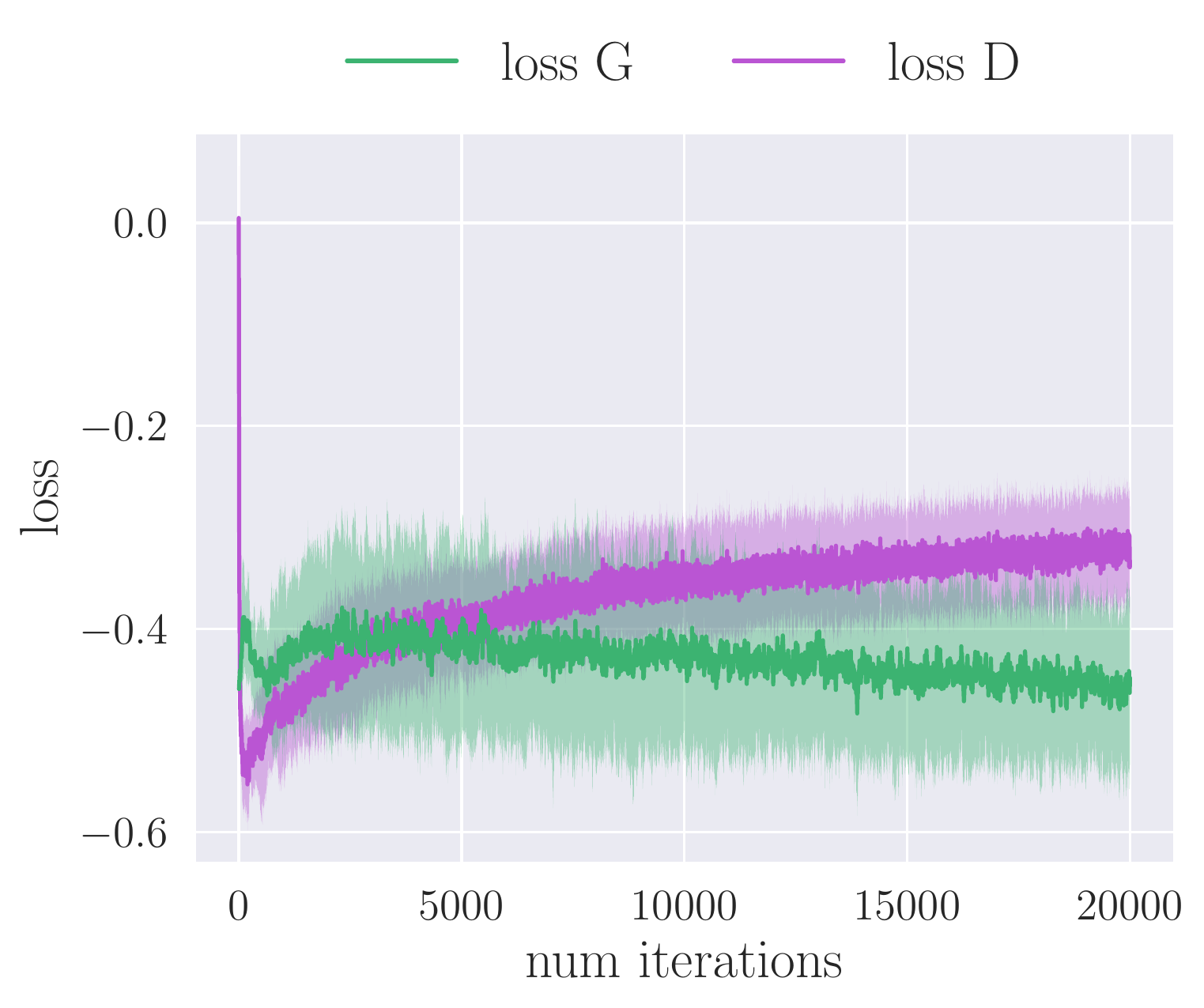}
    \caption{Loss functions for $G$ and $D$ trained on 8 low energy states of~\ref{simple_ising}. $D$ and $G$ both consist of $l = 20$ layers, and $D$ has $N_A = 4$ auxiliary qubits. The loss curves are
averaged over 20 different random initialization. The shaded region in the right panel shows the standard deviation.}
    \label{fig:ising8_loss}
\end{figure}

In this section we train the QGAN on low energy Ising states from a Ising chain of length $N=6$ which has $2^6=64$ total energy eigenstates. As the Ising model Hamiltonian is constrained to a single basis, all of its eigenstates are product states. As training data, we randomly select a subset of $8$ states from the bottom quartile of the energy spectrum. The discriminator is given no information besides these states from which it learns. The loss curves of $G$ and $D$ depicted in Fig.~\ref{fig:ising8_loss} result from training a QGAN of the previously described architecture with $20$ layers in both $G$ and $D$ and $N_A = 4$ auxiliary qubits in $D$.

\begin{figure}
    \centering
    \includegraphics[width=0.9\linewidth]{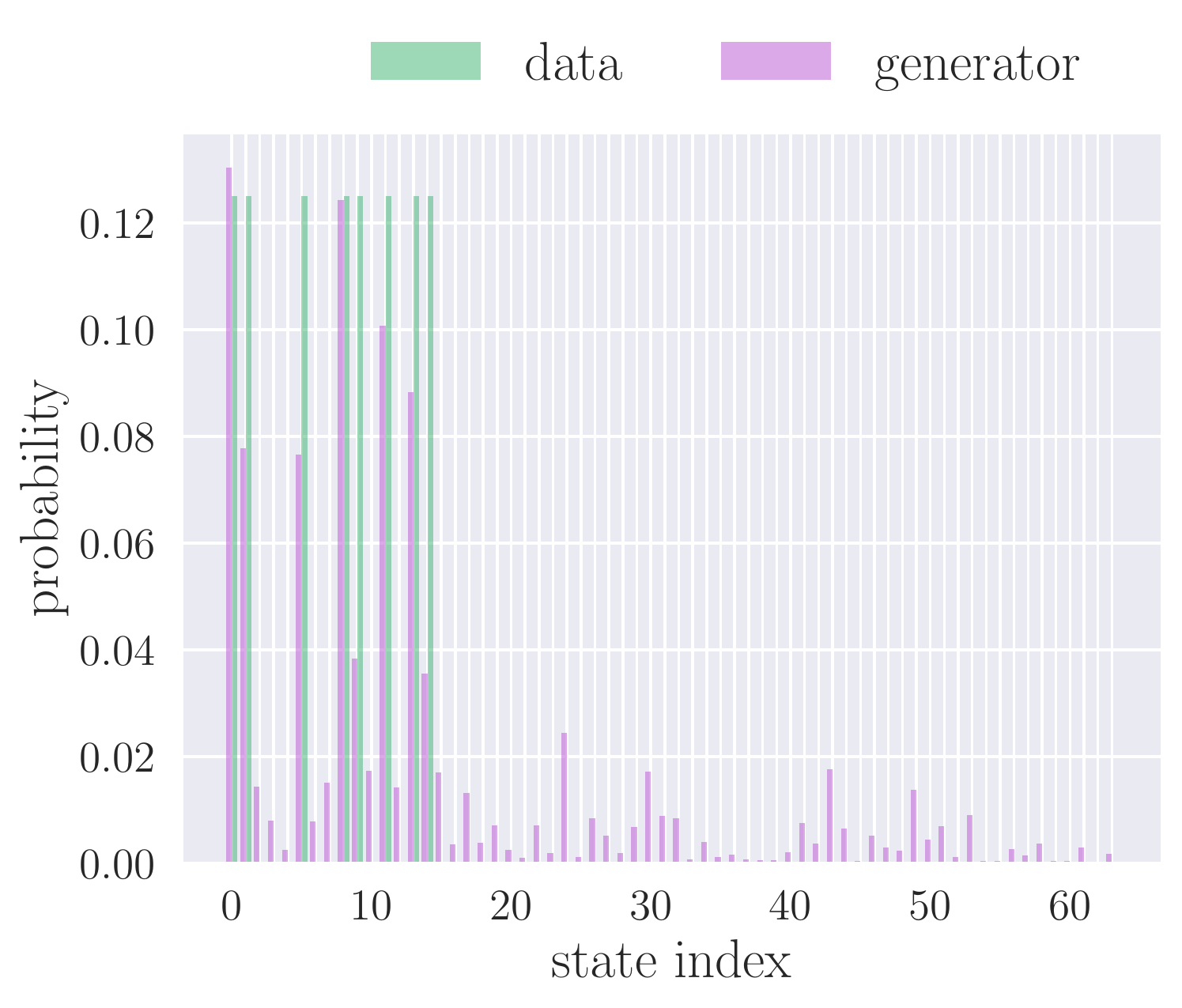}
    \caption{Samples from $G$ trained on 8 low energy states of~\ref{simple_ising}. We draw $10k$ samples from $100$ different noise instances to obtain the histogram of the samples. The x-axis represents the indices of the states ordered by energy. This histogram represents the ouput distribution of the generator for a single training run. Different random parameter initializations lead to similar results.}
    \label{fig:ising8_samples}
\end{figure}

To check the performance of the GAN, we draw samples from the trained generator. A histogram of the resultant states, ordered by energy, is displayed in Fig.~\ref{fig:ising8_samples}. While the average energy of the training states is $-3.0$, the states produced by the generator have an average energy of $-2.6$, indicating that the generator is able to produce states that are below the mean energy of the spectrum ($0$). Since a significant proportion of the states sampled from the fully trained G were repeats of the training data (which is to be expected) we remove the training states from the samples generated by $G$ and the average energy is still negative, at $-0.34$. This quantifies what is visible from the histogram in Fig.~\ref{fig:ising8_samples}. The high energy states to the right are sampled with much lower probability. This suggests that even with a limited number of training samples, $G$ is potentially capable of generalizing and producing novel low energy states at a higher rate than would be expected by chance.

Though in this example the generator was able to produce novel low energy states, the discriminator does not explicitly know that it is training to learn low energy states, and thus may be learning other patterns common to the training set not strictly related to energy, which the generator then learns to reproduce. This problem is somewhat ameliorated by the fact that the $8$ training samples were drawn randomly from a band of the energy spectrum, so that we don't actually teach the generator to produce, say, states of a particular parity. This test serves as a proof-of-concept of using the QGAN to generalize from low energy Ising states. To more rigorously test this capability, larger system sizes and more complicated Ising Hamiltonians should be used, and the required circuit depth and hyperparameters should be studied. We leave these tests for future work. Furthermore, the QGAN tends to converge to a distribution that has zero probabilities for certain states, even those these pertain to the training dataset (index 9 and 14 in Fig.~\ref{fig:ising8_samples}). While clearly some variety of mode collapse, the precise cause of the phenomenon requires further investigation.

\section{Methods}
\label{Sec:Methods}
The numerical simulations in this work are performed with pennylane~\cite{bergholm2018pennylane}, jax~\cite{jax2018github}, TensorLy-Quantum~\cite{patti2021tensorly} and numpy~\cite{harris2020array}.
The code to reproduce this work can be found under~\cite{huembeli_patrick_2022}. Throughout this work we used a Wasserstein protocol where the parameters of the discriminator are updated 5 times more frequently than those of the generator. This variable is referred to as \verb+n_critic = 5+ in the GitLab repository~\cite{huembeli_patrick_2022}. The learning rates for both, the generator and the discriminator are \verb+lrG = lrD = 0.01+. The ADAM optimizer with $\beta_1=0.9$ and $\beta_2=0.999$ is used for all parameter updates~\cite{kingma2014adam}. 

\noindent \textbf{Circuit Architecture for B\&S training:} The discriminators in Figs.~\ref{fig:Data_reuploading_with_noise},~\ref{fig:Toy_Model_no_noise} and~\ref{fig:Toy_Model_with_noise} are all with \verb|depth_D = 20| and $0$ auxiliary qubits. The neural network that is used for Fig.~\ref{fig:Toy_Model_with_noise} has an input dimension of $8$, $8$ hidden nodes and $2^4$ output dimensions. The activation functions are ReLU for the hidden nodes and Tanh for the last layer. The generator for Fig.~\ref{fig:Data_reuploading_with_noise} has \verb|depth_D = 40|.

\noindent \textbf{Circuit Architecture for Ising data:} The discriminator circuit for the training in Figs.~\ref{fig:ising8_samples} and~\ref{fig:ising8_loss} has \verb|depth_D = 40| and \verb|depth_G = 20|.
\section{Discussion and Outlook}
Variational quantum circuits with a hardware efficient architecture are a suitable choice to learn discrete data in a generative adversarial setting. To our knowledge, previous work focused on QGANs with classical discriminators, which require access to the generator's output distribution. This poses a major limitation for the scalability to higher dimensional binary data. We show that a fully coherent quantum architecture is capable of learning discrete data in an adversarial setting and, therefore, does not suffer from this limitation. Furthermore, we show that adding classical noise to the quantum generator is crucial for it to be trained in the adversarial setting. To add noise to the circuit we use the data reuploading architecture. Finally, we benchmark our QGAN architecture on the B\&S dataset and on low energy Ising states and demonstrate that it successfully learns the training data and and achieves aspects of generalization.

With respect to our main goal, the sampling of unknown low energy states from a given Ising Hamiltonian, various open questions remain. For example an in-depth study of QGANs' generalization capability, similar to that done for Born machines~\cite{gili2022quantum} should be performed.
Furthermore, even though our ansatz does not require access to the generator's output distribution $p_G$, its scalability might still be limited by many factors. First, for the given hardware efficient circuit ansatz, barren plateaus will make the training for a large number of qubits impossible~\cite{mcclean2018barren, patti2021entanglement, cerezo2021cost}. As a result, it may be preferable to find a more barren plateau-resistant ansatz for the discriminator --- for example quantum convolutional neural networks \cite{2019NatPh..15.1273C,pesah2021absence} or general dissipative QNNs \cite{sharma2022trainability}.
Second, it is not clear yet how the required depth of the quantum circuit scales with the size of the input data and the number of qubits. The worst case scenario is $\mathcal{O}(4^N)$~\cite{shende2004minimal, rakyta2022approaching}, which would become intractable at scale. Thus, it is be necessary to analyze the scaling of our algorithm's requisite circuit depth and performance with respect to qubit count and data complexity.

In summary, we test our fully-quantum QGAN architecture on a data sets derived from the low-energy states of Ising models with constant, nearest neighbor interactions and no magnetic field. Though the QGAN performed well in our investigations, most QUBO problems encountered in industry sport more complicated Hamiltonians with more intricate connectivity graphs. For instance, these may involve all-to-all coupling with varying magnitude. As a result, it may become important to provide information about the connectivity of the problem to the QGAN as the Hamiltonian grows more complex, as it will become more difficult for the QGAN to infer information about the energy spectrum from a few training states alone. In classical ML this can be accomplished with a conditional GAN~\cite{mirza2014conditional}. Augmenting our QGAN architecture by conditioning on Hamiltonian information would serve as an interesting and useful extension of this work.
\bibliography{main.bib}

\begin{thebibliography}{56}%
\makeatletter
\providecommand \@ifxundefined [1]{%
 \@ifx{#1\undefined}
}%
\providecommand \@ifnum [1]{%
 \ifnum #1\expandafter \@firstoftwo
 \else \expandafter \@secondoftwo
 \fi
}%
\providecommand \@ifx [1]{%
 \ifx #1\expandafter \@firstoftwo
 \else \expandafter \@secondoftwo
 \fi
}%
\providecommand \natexlab [1]{#1}%
\providecommand \enquote  [1]{``#1''}%
\providecommand \bibnamefont  [1]{#1}%
\providecommand \bibfnamefont [1]{#1}%
\providecommand \citenamefont [1]{#1}%
\providecommand \href@noop [0]{\@secondoftwo}%
\providecommand \href [0]{\begingroup \@sanitize@url \@href}%
\providecommand \@href[1]{\@@startlink{#1}\@@href}%
\providecommand \@@href[1]{\endgroup#1\@@endlink}%
\providecommand \@sanitize@url [0]{\catcode `\\12\catcode `\$12\catcode
  `\&12\catcode `\#12\catcode `\^12\catcode `\_12\catcode `\%12\relax}%
\providecommand \@@startlink[1]{}%
\providecommand \@@endlink[0]{}%
\providecommand \url  [0]{\begingroup\@sanitize@url \@url }%
\providecommand \@url [1]{\endgroup\@href {#1}{\urlprefix }}%
\providecommand \urlprefix  [0]{URL }%
\providecommand \Eprint [0]{\href }%
\providecommand \doibase [0]{http://dx.doi.org/}%
\providecommand \selectlanguage [0]{\@gobble}%
\providecommand \bibinfo  [0]{\@secondoftwo}%
\providecommand \bibfield  [0]{\@secondoftwo}%
\providecommand \translation [1]{[#1]}%
\providecommand \BibitemOpen [0]{}%
\providecommand \bibitemStop [0]{}%
\providecommand \bibitemNoStop [0]{.\EOS\space}%
\providecommand \EOS [0]{\spacefactor3000\relax}%
\providecommand \BibitemShut  [1]{\csname bibitem#1\endcsname}%
\let\auto@bib@innerbib\@empty
%</preamble>
\bibitem [{\citenamefont {Xie}\ \emph {et~al.}(2017)\citenamefont {Xie},
  \citenamefont {Girshick}, \citenamefont {Doll{\'a}r}, \citenamefont {Tu},\
  and\ \citenamefont {He}}]{xie2017aggregated}%
  \BibitemOpen
  \bibfield  {author} {\bibinfo {author} {\bibfnamefont {S.}~\bibnamefont
  {Xie}}, \bibinfo {author} {\bibfnamefont {R.}~\bibnamefont {Girshick}},
  \bibinfo {author} {\bibfnamefont {P.}~\bibnamefont {Doll{\'a}r}}, \bibinfo
  {author} {\bibfnamefont {Z.}~\bibnamefont {Tu}}, \ and\ \bibinfo {author}
  {\bibfnamefont {K.}~\bibnamefont {He}},\ }in\ \href@noop {} {\emph {\bibinfo
  {booktitle} {Proceedings of the IEEE conference on computer vision and
  pattern recognition}}}\ (\bibinfo {year} {2017})\ pp.\ \bibinfo {pages}
  {1492--1500}\BibitemShut {NoStop}%
\bibitem [{\citenamefont {Yu}\ \emph {et~al.}(2022)\citenamefont {Yu},
  \citenamefont {Wang}, \citenamefont {Vasudevan}, \citenamefont {Yeung},
  \citenamefont {Seyedhosseini},\ and\ \citenamefont {Wu}}]{yu2022coca}%
  \BibitemOpen
  \bibfield  {author} {\bibinfo {author} {\bibfnamefont {J.}~\bibnamefont
  {Yu}}, \bibinfo {author} {\bibfnamefont {Z.}~\bibnamefont {Wang}}, \bibinfo
  {author} {\bibfnamefont {V.}~\bibnamefont {Vasudevan}}, \bibinfo {author}
  {\bibfnamefont {L.}~\bibnamefont {Yeung}}, \bibinfo {author} {\bibfnamefont
  {M.}~\bibnamefont {Seyedhosseini}}, \ and\ \bibinfo {author} {\bibfnamefont
  {Y.}~\bibnamefont {Wu}},\ }\href@noop {} {\bibfield  {journal} {\bibinfo
  {journal} {arXiv:2205.01917}\ } (\bibinfo {year} {2022})},\ \Eprint
  {http://arxiv.org/abs/2205.01917} {arXiv:2205.01917} \BibitemShut {NoStop}%
\bibitem [{\citenamefont {Ramesh}\ \emph {et~al.}(2021)\citenamefont {Ramesh},
  \citenamefont {Pavlov}, \citenamefont {Goh}, \citenamefont {Gray},
  \citenamefont {Voss}, \citenamefont {Radford}, \citenamefont {Chen},\ and\
  \citenamefont {Sutskever}}]{ramesh2021zero}%
  \BibitemOpen
  \bibfield  {author} {\bibinfo {author} {\bibfnamefont {A.}~\bibnamefont
  {Ramesh}}, \bibinfo {author} {\bibfnamefont {M.}~\bibnamefont {Pavlov}},
  \bibinfo {author} {\bibfnamefont {G.}~\bibnamefont {Goh}}, \bibinfo {author}
  {\bibfnamefont {S.}~\bibnamefont {Gray}}, \bibinfo {author} {\bibfnamefont
  {C.}~\bibnamefont {Voss}}, \bibinfo {author} {\bibfnamefont {A.}~\bibnamefont
  {Radford}}, \bibinfo {author} {\bibfnamefont {M.}~\bibnamefont {Chen}}, \
  and\ \bibinfo {author} {\bibfnamefont {I.}~\bibnamefont {Sutskever}},\ }in\
  \href@noop {} {\emph {\bibinfo {booktitle} {International Conference on
  Machine Learning}}}\ (\bibinfo {organization} {PMLR},\ \bibinfo {year}
  {2021})\ pp.\ \bibinfo {pages} {8821--8831}\BibitemShut {NoStop}%
\bibitem [{\citenamefont {Vasquez}\ and\ \citenamefont
  {Lewis}(2019)}]{vasquez2019melnet}%
  \BibitemOpen
  \bibfield  {author} {\bibinfo {author} {\bibfnamefont {S.}~\bibnamefont
  {Vasquez}}\ and\ \bibinfo {author} {\bibfnamefont {M.}~\bibnamefont
  {Lewis}},\ }\href@noop {} {\bibfield  {journal} {\bibinfo  {journal} {arXiv
  preprint arXiv:1906.01083}\ } (\bibinfo {year} {2019})},\ \Eprint
  {http://arxiv.org/abs/2205.01917} {arXiv:2205.01917} \BibitemShut {NoStop}%
\bibitem [{\citenamefont {Ismail~Fawaz}\ \emph {et~al.}(2019)\citenamefont
  {Ismail~Fawaz}, \citenamefont {Forestier}, \citenamefont {Weber},
  \citenamefont {Idoumghar},\ and\ \citenamefont {Muller}}]{ismail2019deep}%
  \BibitemOpen
  \bibfield  {author} {\bibinfo {author} {\bibfnamefont {H.}~\bibnamefont
  {Ismail~Fawaz}}, \bibinfo {author} {\bibfnamefont {G.}~\bibnamefont
  {Forestier}}, \bibinfo {author} {\bibfnamefont {J.}~\bibnamefont {Weber}},
  \bibinfo {author} {\bibfnamefont {L.}~\bibnamefont {Idoumghar}}, \ and\
  \bibinfo {author} {\bibfnamefont {P.-A.}\ \bibnamefont {Muller}},\
  }\href@noop {} {\bibfield  {journal} {\bibinfo  {journal} {Data mining and
  knowledge discovery}\ }\textbf {\bibinfo {volume} {33}},\ \bibinfo {pages}
  {917} (\bibinfo {year} {2019})}\BibitemShut {NoStop}%
\bibitem [{\citenamefont {Goodfellow}\ \emph {et~al.}(2014)\citenamefont
  {Goodfellow}, \citenamefont {Pouget-Abadie}, \citenamefont {Mirza},
  \citenamefont {Xu}, \citenamefont {Warde-Farley}, \citenamefont {Ozair},
  \citenamefont {Courville},\ and\ \citenamefont
  {Bengio}}]{2014arXiv1406.2661G}%
  \BibitemOpen
  \bibfield  {author} {\bibinfo {author} {\bibfnamefont {I.~J.}\ \bibnamefont
  {Goodfellow}}, \bibinfo {author} {\bibfnamefont {J.}~\bibnamefont
  {Pouget-Abadie}}, \bibinfo {author} {\bibfnamefont {M.}~\bibnamefont
  {Mirza}}, \bibinfo {author} {\bibfnamefont {B.}~\bibnamefont {Xu}}, \bibinfo
  {author} {\bibfnamefont {D.}~\bibnamefont {Warde-Farley}}, \bibinfo {author}
  {\bibfnamefont {S.}~\bibnamefont {Ozair}}, \bibinfo {author} {\bibfnamefont
  {A.~C.}\ \bibnamefont {Courville}}, \ and\ \bibinfo {author} {\bibfnamefont
  {Y.}~\bibnamefont {Bengio}},\ }in\ \href@noop {} {\emph {\bibinfo {booktitle}
  {NIPS}}}\ (\bibinfo {year} {2014})\ \Eprint {http://arxiv.org/abs/1406.2661}
  {arXiv:1406.2661} \BibitemShut {NoStop}%
\bibitem [{\citenamefont {Karras}\ \emph {et~al.}(2019)\citenamefont {Karras},
  \citenamefont {Laine},\ and\ \citenamefont {Aila}}]{2018arXiv181204948K}%
  \BibitemOpen
  \bibfield  {author} {\bibinfo {author} {\bibfnamefont {T.}~\bibnamefont
  {Karras}}, \bibinfo {author} {\bibfnamefont {S.}~\bibnamefont {Laine}}, \
  and\ \bibinfo {author} {\bibfnamefont {T.}~\bibnamefont {Aila}},\ }in\
  \href@noop {} {\emph {\bibinfo {booktitle} {Proceedings of the IEEE/CVF
  conference on computer vision and pattern recognition}}}\ (\bibinfo {year}
  {2019})\ pp.\ \bibinfo {pages} {4401--4410},\ \Eprint
  {http://arxiv.org/abs/1812.04948} {arXiv:1812.04948} \BibitemShut {NoStop}%
\bibitem [{\citenamefont {Karras}\ \emph {et~al.}(2020)\citenamefont {Karras},
  \citenamefont {Laine}, \citenamefont {Aittala}, \citenamefont {Hellsten},
  \citenamefont {Lehtinen},\ and\ \citenamefont {Aila}}]{karras2020analyzing}%
  \BibitemOpen
  \bibfield  {author} {\bibinfo {author} {\bibfnamefont {T.}~\bibnamefont
  {Karras}}, \bibinfo {author} {\bibfnamefont {S.}~\bibnamefont {Laine}},
  \bibinfo {author} {\bibfnamefont {M.}~\bibnamefont {Aittala}}, \bibinfo
  {author} {\bibfnamefont {J.}~\bibnamefont {Hellsten}}, \bibinfo {author}
  {\bibfnamefont {J.}~\bibnamefont {Lehtinen}}, \ and\ \bibinfo {author}
  {\bibfnamefont {T.}~\bibnamefont {Aila}},\ }in\ \href@noop {} {\emph
  {\bibinfo {booktitle} {Proceedings of the IEEE/CVF conference on computer
  vision and pattern recognition}}}\ (\bibinfo {year} {2020})\ pp.\ \bibinfo
  {pages} {8110--8119},\ \Eprint {http://arxiv.org/abs/1912.04958}
  {arXiv:1912.04958} \BibitemShut {NoStop}%
\bibitem [{\citenamefont {Karras}\ \emph {et~al.}(2021)\citenamefont {Karras},
  \citenamefont {Aittala}, \citenamefont {Laine}, \citenamefont
  {H{\"a}rk{\"o}nen}, \citenamefont {Hellsten}, \citenamefont {Lehtinen},\ and\
  \citenamefont {Aila}}]{karras2021alias}%
  \BibitemOpen
  \bibfield  {author} {\bibinfo {author} {\bibfnamefont {T.}~\bibnamefont
  {Karras}}, \bibinfo {author} {\bibfnamefont {M.}~\bibnamefont {Aittala}},
  \bibinfo {author} {\bibfnamefont {S.}~\bibnamefont {Laine}}, \bibinfo
  {author} {\bibfnamefont {E.}~\bibnamefont {H{\"a}rk{\"o}nen}}, \bibinfo
  {author} {\bibfnamefont {J.}~\bibnamefont {Hellsten}}, \bibinfo {author}
  {\bibfnamefont {J.}~\bibnamefont {Lehtinen}}, \ and\ \bibinfo {author}
  {\bibfnamefont {T.}~\bibnamefont {Aila}},\ }\href@noop {} {\bibfield
  {journal} {\bibinfo  {journal} {Advances in Neural Information Processing
  Systems}\ }\textbf {\bibinfo {volume} {34}},\ \bibinfo {pages} {852}
  (\bibinfo {year} {2021})},\ \Eprint {http://arxiv.org/abs/2106.12423}
  {arXiv:2106.12423} \BibitemShut {NoStop}%
\bibitem [{\citenamefont {Du}\ \emph {et~al.}(2020)\citenamefont {Du},
  \citenamefont {Hsieh}, \citenamefont {Liu},\ and\ \citenamefont
  {Tao}}]{2018arXiv181011922D}%
  \BibitemOpen
  \bibfield  {author} {\bibinfo {author} {\bibfnamefont {Y.}~\bibnamefont
  {Du}}, \bibinfo {author} {\bibfnamefont {M.-H.}\ \bibnamefont {Hsieh}},
  \bibinfo {author} {\bibfnamefont {T.}~\bibnamefont {Liu}}, \ and\ \bibinfo
  {author} {\bibfnamefont {D.}~\bibnamefont {Tao}},\ }\href {\doibase
  10.1103/PhysRevResearch.2.033125} {\bibfield  {journal} {\bibinfo  {journal}
  {Phys. Rev. Research}\ }\textbf {\bibinfo {volume} {2}},\ \bibinfo {pages}
  {033125} (\bibinfo {year} {2020})}\BibitemShut {NoStop}%
\bibitem [{\citenamefont {Schuld}\ and\ \citenamefont
  {Petruccione}(2021)}]{schuld2021machine}%
  \BibitemOpen
  \bibfield  {author} {\bibinfo {author} {\bibfnamefont {M.}~\bibnamefont
  {Schuld}}\ and\ \bibinfo {author} {\bibfnamefont {F.}~\bibnamefont
  {Petruccione}},\ }\href@noop {} {\emph {\bibinfo {title} {Machine learning
  with quantum computers}}}\ (\bibinfo  {publisher} {Springer},\ \bibinfo
  {year} {2021})\BibitemShut {NoStop}%
\bibitem [{\citenamefont {Dallaire-Demers}\ and\ \citenamefont
  {Killoran}(2018)}]{dallaire2018quantum}%
  \BibitemOpen
  \bibfield  {author} {\bibinfo {author} {\bibfnamefont {P.-L.}\ \bibnamefont
  {Dallaire-Demers}}\ and\ \bibinfo {author} {\bibfnamefont {N.}~\bibnamefont
  {Killoran}},\ }\href@noop {} {\bibfield  {journal} {\bibinfo  {journal}
  {Physical Review A}\ }\textbf {\bibinfo {volume} {98}},\ \bibinfo {pages}
  {012324} (\bibinfo {year} {2018})},\ \Eprint
  {http://arxiv.org/abs/1804.08641} {arXiv:1804.08641} \BibitemShut {NoStop}%
\bibitem [{\citenamefont {Zoufal}\ \emph {et~al.}(2019)\citenamefont {Zoufal},
  \citenamefont {Lucchi},\ and\ \citenamefont {Woerner}}]{zoufal2019quantum}%
  \BibitemOpen
  \bibfield  {author} {\bibinfo {author} {\bibfnamefont {C.}~\bibnamefont
  {Zoufal}}, \bibinfo {author} {\bibfnamefont {A.}~\bibnamefont {Lucchi}}, \
  and\ \bibinfo {author} {\bibfnamefont {S.}~\bibnamefont {Woerner}},\
  }\href@noop {} {\bibfield  {journal} {\bibinfo  {journal} {npj Quantum
  Information}\ }\textbf {\bibinfo {volume} {5}},\ \bibinfo {pages} {1}
  (\bibinfo {year} {2019})},\ \Eprint {http://arxiv.org/abs/1904.00043}
  {arXiv:1904.00043} \BibitemShut {NoStop}%
\bibitem [{\citenamefont {Zeng}\ \emph {et~al.}(2019)\citenamefont {Zeng},
  \citenamefont {Wu}, \citenamefont {Liu}, \citenamefont {Wang},\ and\
  \citenamefont {Hu}}]{zeng2019learning}%
  \BibitemOpen
  \bibfield  {author} {\bibinfo {author} {\bibfnamefont {J.}~\bibnamefont
  {Zeng}}, \bibinfo {author} {\bibfnamefont {Y.}~\bibnamefont {Wu}}, \bibinfo
  {author} {\bibfnamefont {J.-G.}\ \bibnamefont {Liu}}, \bibinfo {author}
  {\bibfnamefont {L.}~\bibnamefont {Wang}}, \ and\ \bibinfo {author}
  {\bibfnamefont {J.}~\bibnamefont {Hu}},\ }\href@noop {} {\bibfield  {journal}
  {\bibinfo  {journal} {Physical Review A}\ }\textbf {\bibinfo {volume} {99}},\
  \bibinfo {pages} {052306} (\bibinfo {year} {2019})},\ \Eprint
  {http://arxiv.org/abs/1808.03425} {arXiv:1808.03425} \BibitemShut {NoStop}%
\bibitem [{\citenamefont {Situ}\ \emph {et~al.}(2020)\citenamefont {Situ},
  \citenamefont {He}, \citenamefont {Wang}, \citenamefont {Li},\ and\
  \citenamefont {Zheng}}]{situ2020quantum}%
  \BibitemOpen
  \bibfield  {author} {\bibinfo {author} {\bibfnamefont {H.}~\bibnamefont
  {Situ}}, \bibinfo {author} {\bibfnamefont {Z.}~\bibnamefont {He}}, \bibinfo
  {author} {\bibfnamefont {Y.}~\bibnamefont {Wang}}, \bibinfo {author}
  {\bibfnamefont {L.}~\bibnamefont {Li}}, \ and\ \bibinfo {author}
  {\bibfnamefont {S.}~\bibnamefont {Zheng}},\ }\href@noop {} {\bibfield
  {journal} {\bibinfo  {journal} {Information Sciences}\ }\textbf {\bibinfo
  {volume} {538}},\ \bibinfo {pages} {193} (\bibinfo {year} {2020})},\ \Eprint
  {http://arxiv.org/abs/1807.01235} {arXiv:1807.01235} \BibitemShut {NoStop}%
\bibitem [{\citenamefont {Huang}\ \emph {et~al.}(2021)\citenamefont {Huang},
  \citenamefont {Du}, \citenamefont {Gong}, \citenamefont {Zhao}, \citenamefont
  {Wu}, \citenamefont {Wang}, \citenamefont {Li}, \citenamefont {Liang},
  \citenamefont {Lin}, \citenamefont {Xu} \emph
  {et~al.}}]{huang2021experimental}%
  \BibitemOpen
  \bibfield  {author} {\bibinfo {author} {\bibfnamefont {H.-L.}\ \bibnamefont
  {Huang}}, \bibinfo {author} {\bibfnamefont {Y.}~\bibnamefont {Du}}, \bibinfo
  {author} {\bibfnamefont {M.}~\bibnamefont {Gong}}, \bibinfo {author}
  {\bibfnamefont {Y.}~\bibnamefont {Zhao}}, \bibinfo {author} {\bibfnamefont
  {Y.}~\bibnamefont {Wu}}, \bibinfo {author} {\bibfnamefont {C.}~\bibnamefont
  {Wang}}, \bibinfo {author} {\bibfnamefont {S.}~\bibnamefont {Li}}, \bibinfo
  {author} {\bibfnamefont {F.}~\bibnamefont {Liang}}, \bibinfo {author}
  {\bibfnamefont {J.}~\bibnamefont {Lin}}, \bibinfo {author} {\bibfnamefont
  {Y.}~\bibnamefont {Xu}},  \emph {et~al.},\ }\href@noop {} {\bibfield
  {journal} {\bibinfo  {journal} {Physical Review Applied}\ }\textbf {\bibinfo
  {volume} {16}},\ \bibinfo {pages} {024051} (\bibinfo {year} {2021})},\
  \Eprint {http://arxiv.org/abs/2010.06201} {arXiv:2010.06201} \BibitemShut
  {NoStop}%
\bibitem [{\citenamefont {Stein}\ \emph {et~al.}(2021)\citenamefont {Stein},
  \citenamefont {Baheri}, \citenamefont {Chen}, \citenamefont {Mao},
  \citenamefont {Guan}, \citenamefont {Li}, \citenamefont {Fang},\ and\
  \citenamefont {Xu}}]{stein2021qugan}%
  \BibitemOpen
  \bibfield  {author} {\bibinfo {author} {\bibfnamefont {S.~A.}\ \bibnamefont
  {Stein}}, \bibinfo {author} {\bibfnamefont {B.}~\bibnamefont {Baheri}},
  \bibinfo {author} {\bibfnamefont {D.}~\bibnamefont {Chen}}, \bibinfo {author}
  {\bibfnamefont {Y.}~\bibnamefont {Mao}}, \bibinfo {author} {\bibfnamefont
  {Q.}~\bibnamefont {Guan}}, \bibinfo {author} {\bibfnamefont {A.}~\bibnamefont
  {Li}}, \bibinfo {author} {\bibfnamefont {B.}~\bibnamefont {Fang}}, \ and\
  \bibinfo {author} {\bibfnamefont {S.}~\bibnamefont {Xu}},\ }in\ \href@noop {}
  {\emph {\bibinfo {booktitle} {2021 IEEE International Conference on Quantum
  Computing and Engineering (QCE)}}}\ (\bibinfo {organization} {IEEE},\
  \bibinfo {year} {2021})\ pp.\ \bibinfo {pages} {71--81},\ \Eprint
  {http://arxiv.org/abs/2010.09036} {arXiv:2010.09036} \BibitemShut {NoStop}%
\bibitem [{\citenamefont {Beer}\ and\ \citenamefont
  {M{\"u}ller}(2021)}]{beer2021dissipative}%
  \BibitemOpen
  \bibfield  {author} {\bibinfo {author} {\bibfnamefont {K.}~\bibnamefont
  {Beer}}\ and\ \bibinfo {author} {\bibfnamefont {G.}~\bibnamefont
  {M{\"u}ller}},\ }\href@noop {} {\bibfield  {journal} {\bibinfo  {journal}
  {arXiv: 2112.06088}\ } (\bibinfo {year} {2021})},\ \Eprint
  {http://arxiv.org/abs/2112.06088} {arXiv:2112.06088} \BibitemShut {NoStop}%
\bibitem [{\citenamefont {Niu}\ \emph {et~al.}(2022)\citenamefont {Niu},
  \citenamefont {Zlokapa}, \citenamefont {Broughton}, \citenamefont {Boixo},
  \citenamefont {Mohseni}, \citenamefont {Smelyanskyi},\ and\ \citenamefont
  {Neven}}]{niu2022entangling}%
  \BibitemOpen
  \bibfield  {author} {\bibinfo {author} {\bibfnamefont {M.~Y.}\ \bibnamefont
  {Niu}}, \bibinfo {author} {\bibfnamefont {A.}~\bibnamefont {Zlokapa}},
  \bibinfo {author} {\bibfnamefont {M.}~\bibnamefont {Broughton}}, \bibinfo
  {author} {\bibfnamefont {S.}~\bibnamefont {Boixo}}, \bibinfo {author}
  {\bibfnamefont {M.}~\bibnamefont {Mohseni}}, \bibinfo {author} {\bibfnamefont
  {V.}~\bibnamefont {Smelyanskyi}}, \ and\ \bibinfo {author} {\bibfnamefont
  {H.}~\bibnamefont {Neven}},\ }\href@noop {} {\bibfield  {journal} {\bibinfo
  {journal} {Physical Review Letters}\ }\textbf {\bibinfo {volume} {128}},\
  \bibinfo {pages} {220505} (\bibinfo {year} {2022})},\ \Eprint
  {http://arxiv.org/abs/2105.00080} {arXiv:2105.00080} \BibitemShut {NoStop}%
\bibitem [{\citenamefont {Borras}\ \emph {et~al.}(2022)\citenamefont {Borras},
  \citenamefont {Chang}, \citenamefont {Funcke}, \citenamefont {Grossi},
  \citenamefont {Hartung}, \citenamefont {Jansen}, \citenamefont {Kruecker},
  \citenamefont {K{\"u}hn}, \citenamefont {Rehm}, \citenamefont
  {T{\"u}ys{\"u}z} \emph {et~al.}}]{borras2022impact}%
  \BibitemOpen
  \bibfield  {author} {\bibinfo {author} {\bibfnamefont {K.}~\bibnamefont
  {Borras}}, \bibinfo {author} {\bibfnamefont {S.~Y.}\ \bibnamefont {Chang}},
  \bibinfo {author} {\bibfnamefont {L.}~\bibnamefont {Funcke}}, \bibinfo
  {author} {\bibfnamefont {M.}~\bibnamefont {Grossi}}, \bibinfo {author}
  {\bibfnamefont {T.}~\bibnamefont {Hartung}}, \bibinfo {author} {\bibfnamefont
  {K.}~\bibnamefont {Jansen}}, \bibinfo {author} {\bibfnamefont
  {D.}~\bibnamefont {Kruecker}}, \bibinfo {author} {\bibfnamefont
  {S.}~\bibnamefont {K{\"u}hn}}, \bibinfo {author} {\bibfnamefont
  {F.}~\bibnamefont {Rehm}}, \bibinfo {author} {\bibfnamefont {C.}~\bibnamefont
  {T{\"u}ys{\"u}z}},  \emph {et~al.},\ }\href@noop {} {\bibfield  {journal}
  {\bibinfo  {journal} {arXiv:2203.01007}\ } (\bibinfo {year} {2022})},\
  \Eprint {http://arxiv.org/abs/2203.01007} {arXiv:2203.01007} \BibitemShut
  {NoStop}%
\bibitem [{\citenamefont {Chang}\ \emph {et~al.}(2022)\citenamefont {Chang},
  \citenamefont {Agnew}, \citenamefont {Combarro}, \citenamefont {Grossi},
  \citenamefont {Herbert},\ and\ \citenamefont
  {Vallecorsa}}]{chang2022running}%
  \BibitemOpen
  \bibfield  {author} {\bibinfo {author} {\bibfnamefont {S.~Y.}\ \bibnamefont
  {Chang}}, \bibinfo {author} {\bibfnamefont {E.}~\bibnamefont {Agnew}},
  \bibinfo {author} {\bibfnamefont {E.~F.}\ \bibnamefont {Combarro}}, \bibinfo
  {author} {\bibfnamefont {M.}~\bibnamefont {Grossi}}, \bibinfo {author}
  {\bibfnamefont {S.}~\bibnamefont {Herbert}}, \ and\ \bibinfo {author}
  {\bibfnamefont {S.}~\bibnamefont {Vallecorsa}},\ }\href@noop {} {\bibfield
  {journal} {\bibinfo  {journal} {arXiv:2205.15003}\ } (\bibinfo {year}
  {2022})},\ \Eprint {http://arxiv.org/abs/2205.15003} {arXiv:2205.15003}
  \BibitemShut {NoStop}%
\bibitem [{\citenamefont {Rudolph}\ \emph {et~al.}(2022)\citenamefont
  {Rudolph}, \citenamefont {Toussaint}, \citenamefont {Katabarwa},
  \citenamefont {Johri}, \citenamefont {Peropadre},\ and\ \citenamefont
  {Perdomo-Ortiz}}]{rudolph2022generation}%
  \BibitemOpen
  \bibfield  {author} {\bibinfo {author} {\bibfnamefont {M.~S.}\ \bibnamefont
  {Rudolph}}, \bibinfo {author} {\bibfnamefont {N.~B.}\ \bibnamefont
  {Toussaint}}, \bibinfo {author} {\bibfnamefont {A.}~\bibnamefont
  {Katabarwa}}, \bibinfo {author} {\bibfnamefont {S.}~\bibnamefont {Johri}},
  \bibinfo {author} {\bibfnamefont {B.}~\bibnamefont {Peropadre}}, \ and\
  \bibinfo {author} {\bibfnamefont {A.}~\bibnamefont {Perdomo-Ortiz}},\
  }\href@noop {} {\bibfield  {journal} {\bibinfo  {journal} {Physical Review
  X}\ }\textbf {\bibinfo {volume} {12}},\ \bibinfo {pages} {031010} (\bibinfo
  {year} {2022})},\ \Eprint {http://arxiv.org/abs/2012.03924}
  {arXiv:2012.03924} \BibitemShut {NoStop}%
\bibitem [{\citenamefont {Rumelhart}\ \emph {et~al.}(1988)\citenamefont
  {Rumelhart}, \citenamefont {Hinton},\ and\ \citenamefont
  {Williams}}]{backprop}%
  \BibitemOpen
  \bibfield  {author} {\bibinfo {author} {\bibfnamefont {D.~E.}\ \bibnamefont
  {Rumelhart}}, \bibinfo {author} {\bibfnamefont {G.~E.}\ \bibnamefont
  {Hinton}}, \ and\ \bibinfo {author} {\bibfnamefont {R.~J.}\ \bibnamefont
  {Williams}},\ }\enquote {\bibinfo {title} {Learning representations by
  back-propagating errors},}\ in\ \href@noop {} {\emph {\bibinfo {booktitle}
  {Neurocomputing: Foundations of Research}}}\ (\bibinfo  {publisher} {MIT
  Press},\ \bibinfo {address} {Cambridge, MA, USA},\ \bibinfo {year} {1988})\
  p.\ \bibinfo {pages} {696–699}\BibitemShut {NoStop}%
\bibitem [{\citenamefont {{Alcazar}}\ \emph {et~al.}(2021)\citenamefont
  {{Alcazar}}, \citenamefont {{Ghazi Vakili}}, \citenamefont {{Kalayci}},\ and\
  \citenamefont {{Perdomo-Ortiz}}}]{2021arXiv210106250A}%
  \BibitemOpen
  \bibfield  {author} {\bibinfo {author} {\bibfnamefont {J.}~\bibnamefont
  {{Alcazar}}}, \bibinfo {author} {\bibfnamefont {M.}~\bibnamefont {{Ghazi
  Vakili}}}, \bibinfo {author} {\bibfnamefont {C.~B.}\ \bibnamefont
  {{Kalayci}}}, \ and\ \bibinfo {author} {\bibfnamefont {A.}~\bibnamefont
  {{Perdomo-Ortiz}}},\ }\href@noop {} {\bibfield  {journal} {\bibinfo
  {journal} {arXiv e-prints}\ } (\bibinfo {year} {2021})},\ \Eprint
  {http://arxiv.org/abs/2101.06250} {arXiv:2101.06250} \BibitemShut {NoStop}%
\bibitem [{\citenamefont {{Gili}}\ \emph {et~al.}(2022)\citenamefont {{Gili}},
  \citenamefont {{Hibat-Allah}}, \citenamefont {{Mauri}}, \citenamefont
  {{Ballance}},\ and\ \citenamefont {{Perdomo-Ortiz}}}]{2022arXiv220713645G}%
  \BibitemOpen
  \bibfield  {author} {\bibinfo {author} {\bibfnamefont {K.}~\bibnamefont
  {{Gili}}}, \bibinfo {author} {\bibfnamefont {M.}~\bibnamefont
  {{Hibat-Allah}}}, \bibinfo {author} {\bibfnamefont {M.}~\bibnamefont
  {{Mauri}}}, \bibinfo {author} {\bibfnamefont {C.}~\bibnamefont {{Ballance}}},
  \ and\ \bibinfo {author} {\bibfnamefont {A.}~\bibnamefont
  {{Perdomo-Ortiz}}},\ }\href@noop {} {\bibfield  {journal} {\bibinfo
  {journal} {arXiv e-prints}\ } (\bibinfo {year} {2022})},\ \Eprint
  {http://arxiv.org/abs/2207.13645} {arXiv:2207.13645} \BibitemShut {NoStop}%
\bibitem [{\citenamefont {Gili}\ \emph
  {et~al.}(2022{\natexlab{a}})\citenamefont {Gili}, \citenamefont {Mauri},\
  and\ \citenamefont {Perdomo-Ortiz}}]{gili2022evaluating}%
  \BibitemOpen
  \bibfield  {author} {\bibinfo {author} {\bibfnamefont {K.}~\bibnamefont
  {Gili}}, \bibinfo {author} {\bibfnamefont {M.}~\bibnamefont {Mauri}}, \ and\
  \bibinfo {author} {\bibfnamefont {A.}~\bibnamefont {Perdomo-Ortiz}},\
  }\href@noop {} {\bibfield  {journal} {\bibinfo  {journal} {arXiv:2201.08770}\
  } (\bibinfo {year} {2022}{\natexlab{a}})},\ \Eprint
  {http://arxiv.org/abs/2201.08770} {arXiv:2201.08770} \BibitemShut {NoStop}%
\bibitem [{\citenamefont {Banchi}\ \emph {et~al.}(2021)\citenamefont {Banchi},
  \citenamefont {Pereira},\ and\ \citenamefont
  {Pirandola}}]{banchi2021generalization}%
  \BibitemOpen
  \bibfield  {author} {\bibinfo {author} {\bibfnamefont {L.}~\bibnamefont
  {Banchi}}, \bibinfo {author} {\bibfnamefont {J.}~\bibnamefont {Pereira}}, \
  and\ \bibinfo {author} {\bibfnamefont {S.}~\bibnamefont {Pirandola}},\
  }\href@noop {} {\bibfield  {journal} {\bibinfo  {journal} {PRX Quantum}\
  }\textbf {\bibinfo {volume} {2}},\ \bibinfo {pages} {040321} (\bibinfo {year}
  {2021})},\ \Eprint {http://arxiv.org/abs/2102.08991} {arXiv:2102.08991}
  \BibitemShut {NoStop}%
\bibitem [{\citenamefont {Arjovsky}\ \emph {et~al.}(2017)\citenamefont
  {Arjovsky}, \citenamefont {Chintala},\ and\ \citenamefont
  {Bottou}}]{arjovsky2017wasserstein}%
  \BibitemOpen
  \bibfield  {author} {\bibinfo {author} {\bibfnamefont {M.}~\bibnamefont
  {Arjovsky}}, \bibinfo {author} {\bibfnamefont {S.}~\bibnamefont {Chintala}},
  \ and\ \bibinfo {author} {\bibfnamefont {L.}~\bibnamefont {Bottou}},\ }in\
  \href@noop {} {\emph {\bibinfo {booktitle} {International conference on
  machine learning}}}\ (\bibinfo {organization} {PMLR},\ \bibinfo {year}
  {2017})\ pp.\ \bibinfo {pages} {214--223},\ \Eprint
  {http://arxiv.org/abs/1701.07875} {arXiv:1701.07875} \BibitemShut {NoStop}%
\bibitem [{\citenamefont {Kiani}\ \emph {et~al.}(2021)\citenamefont {Kiani},
  \citenamefont {De~Palma}, \citenamefont {Marvian}, \citenamefont {Liu},\ and\
  \citenamefont {Lloyd}}]{kiani2021quantum}%
  \BibitemOpen
  \bibfield  {author} {\bibinfo {author} {\bibfnamefont {B.~T.}\ \bibnamefont
  {Kiani}}, \bibinfo {author} {\bibfnamefont {G.}~\bibnamefont {De~Palma}},
  \bibinfo {author} {\bibfnamefont {M.}~\bibnamefont {Marvian}}, \bibinfo
  {author} {\bibfnamefont {Z.-W.}\ \bibnamefont {Liu}}, \ and\ \bibinfo
  {author} {\bibfnamefont {S.}~\bibnamefont {Lloyd}},\ }\href@noop {}
  {\bibfield  {journal} {\bibinfo  {journal} {arXiv:2101.03037}\ } (\bibinfo
  {year} {2021})},\ \Eprint {http://arxiv.org/abs/2101.03037}
  {arXiv:2101.03037} \BibitemShut {NoStop}%
\bibitem [{\citenamefont {Benedetti}\ \emph {et~al.}(2019)\citenamefont
  {Benedetti}, \citenamefont {Garcia-Pintos}, \citenamefont {Perdomo},
  \citenamefont {Leyton-Ortega}, \citenamefont {Nam},\ and\ \citenamefont
  {Perdomo-Ortiz}}]{benedetti2019generative}%
  \BibitemOpen
  \bibfield  {author} {\bibinfo {author} {\bibfnamefont {M.}~\bibnamefont
  {Benedetti}}, \bibinfo {author} {\bibfnamefont {D.}~\bibnamefont
  {Garcia-Pintos}}, \bibinfo {author} {\bibfnamefont {O.}~\bibnamefont
  {Perdomo}}, \bibinfo {author} {\bibfnamefont {V.}~\bibnamefont
  {Leyton-Ortega}}, \bibinfo {author} {\bibfnamefont {Y.}~\bibnamefont {Nam}},
  \ and\ \bibinfo {author} {\bibfnamefont {A.}~\bibnamefont {Perdomo-Ortiz}},\
  }\href@noop {} {\bibfield  {journal} {\bibinfo  {journal} {npj Quantum
  Information}\ }\textbf {\bibinfo {volume} {5}},\ \bibinfo {pages} {1}
  (\bibinfo {year} {2019})},\ \Eprint {http://arxiv.org/abs/1801.07686}
  {arXiv:1801.07686} \BibitemShut {NoStop}%
\bibitem [{\citenamefont {Lucas}(2014)}]{lucas2014ising}%
  \BibitemOpen
  \bibfield  {author} {\bibinfo {author} {\bibfnamefont {A.}~\bibnamefont
  {Lucas}},\ }\href@noop {} {\bibfield  {journal} {\bibinfo  {journal}
  {Frontiers in physics}\ ,\ \bibinfo {pages} {5}} (\bibinfo {year} {2014})},\
  \Eprint {http://arxiv.org/abs/1302.5843} {arXiv:1302.5843} \BibitemShut
  {NoStop}%
\bibitem [{\citenamefont {Mulligan}\ \emph {et~al.}(2020)\citenamefont
  {Mulligan}, \citenamefont {Melo}, \citenamefont {Merritt}, \citenamefont
  {Slocum}, \citenamefont {Weitzner}, \citenamefont {Watkins}, \citenamefont
  {Renfrew}, \citenamefont {Pelissier}, \citenamefont {Arora},\ and\
  \citenamefont {Bonneau}}]{mulligan2020designing}%
  \BibitemOpen
  \bibfield  {author} {\bibinfo {author} {\bibfnamefont {V.~K.}\ \bibnamefont
  {Mulligan}}, \bibinfo {author} {\bibfnamefont {H.}~\bibnamefont {Melo}},
  \bibinfo {author} {\bibfnamefont {H.~I.}\ \bibnamefont {Merritt}}, \bibinfo
  {author} {\bibfnamefont {S.}~\bibnamefont {Slocum}}, \bibinfo {author}
  {\bibfnamefont {B.~D.}\ \bibnamefont {Weitzner}}, \bibinfo {author}
  {\bibfnamefont {A.~M.}\ \bibnamefont {Watkins}}, \bibinfo {author}
  {\bibfnamefont {P.~D.}\ \bibnamefont {Renfrew}}, \bibinfo {author}
  {\bibfnamefont {C.}~\bibnamefont {Pelissier}}, \bibinfo {author}
  {\bibfnamefont {P.~S.}\ \bibnamefont {Arora}}, \ and\ \bibinfo {author}
  {\bibfnamefont {R.}~\bibnamefont {Bonneau}},\ }\href {\doibase
  https://doi.org/10.1101/752485} {\bibfield  {journal} {\bibinfo  {journal}
  {BioRxiv}\ ,\ \bibinfo {pages} {752485}} (\bibinfo {year}
  {2020})}\BibitemShut {NoStop}%
\bibitem [{\citenamefont {Jain}(2021)}]{jain2021solving}%
  \BibitemOpen
  \bibfield  {author} {\bibinfo {author} {\bibfnamefont {S.}~\bibnamefont
  {Jain}},\ }\href {\doibase https://doi.org/10.3389/fphy.2021.7607831}
  {\bibfield  {journal} {\bibinfo  {journal} {Frontiers in Physics}\ ,\
  \bibinfo {pages} {646}} (\bibinfo {year} {2021})}\BibitemShut {NoStop}%
\bibitem [{\citenamefont {Herman}\ \emph {et~al.}(2022)\citenamefont {Herman},
  \citenamefont {Googin}, \citenamefont {Liu}, \citenamefont {Galda},
  \citenamefont {Safro}, \citenamefont {Sun}, \citenamefont {Pistoia},\ and\
  \citenamefont {Alexeev}}]{herman2022survey}%
  \BibitemOpen
  \bibfield  {author} {\bibinfo {author} {\bibfnamefont {D.}~\bibnamefont
  {Herman}}, \bibinfo {author} {\bibfnamefont {C.}~\bibnamefont {Googin}},
  \bibinfo {author} {\bibfnamefont {X.}~\bibnamefont {Liu}}, \bibinfo {author}
  {\bibfnamefont {A.}~\bibnamefont {Galda}}, \bibinfo {author} {\bibfnamefont
  {I.}~\bibnamefont {Safro}}, \bibinfo {author} {\bibfnamefont
  {Y.}~\bibnamefont {Sun}}, \bibinfo {author} {\bibfnamefont {M.}~\bibnamefont
  {Pistoia}}, \ and\ \bibinfo {author} {\bibfnamefont {Y.}~\bibnamefont
  {Alexeev}},\ }\href@noop {} {\bibfield  {journal} {\bibinfo  {journal}
  {arXiv:2201.02773}\ } (\bibinfo {year} {2022})},\ \Eprint
  {http://arxiv.org/abs/2201.02773} {arXiv:2201.02773} \BibitemShut {NoStop}%
\bibitem [{\citenamefont {Dean}\ and\ \citenamefont
  {Lewis}(1999)}]{dean1999molecular}%
  \BibitemOpen
  \bibfield  {author} {\bibinfo {author} {\bibfnamefont {P.~M.}\ \bibnamefont
  {Dean}}\ and\ \bibinfo {author} {\bibfnamefont {R.~A.}\ \bibnamefont
  {Lewis}},\ }\href@noop {} {\emph {\bibinfo {title} {Molecular diversity in
  drug design}}}\ (\bibinfo  {publisher} {Springer},\ \bibinfo {year}
  {1999})\BibitemShut {NoStop}%
\bibitem [{\citenamefont {Gorse}\ \emph {et~al.}(1999)\citenamefont {Gorse},
  \citenamefont {Rees}, \citenamefont {Kaczorek},\ and\ \citenamefont
  {Lahana}}]{gorse_molecular_1999}%
  \BibitemOpen
  \bibfield  {author} {\bibinfo {author} {\bibfnamefont {D.}~\bibnamefont
  {Gorse}}, \bibinfo {author} {\bibfnamefont {A.}~\bibnamefont {Rees}},
  \bibinfo {author} {\bibfnamefont {M.}~\bibnamefont {Kaczorek}}, \ and\
  \bibinfo {author} {\bibfnamefont {R.}~\bibnamefont {Lahana}},\ }\href
  {\doibase 10.1016/s1359-6446(99)01334-3} {\bibfield  {journal} {\bibinfo
  {journal} {Drug Discovery Today}\ }\textbf {\bibinfo {volume} {4}},\ \bibinfo
  {pages} {257} (\bibinfo {year} {1999})}\BibitemShut {NoStop}%
\bibitem [{\citenamefont {Galloway}\ \emph {et~al.}(2010)\citenamefont
  {Galloway}, \citenamefont {Isidro-Llobet},\ and\ \citenamefont
  {Spring}}]{galloway2010diversity}%
  \BibitemOpen
  \bibfield  {author} {\bibinfo {author} {\bibfnamefont {W.~R.}\ \bibnamefont
  {Galloway}}, \bibinfo {author} {\bibfnamefont {A.}~\bibnamefont
  {Isidro-Llobet}}, \ and\ \bibinfo {author} {\bibfnamefont {D.~R.}\
  \bibnamefont {Spring}},\ }\href@noop {} {\bibfield  {journal} {\bibinfo
  {journal} {Nature communications}\ }\textbf {\bibinfo {volume} {1}},\
  \bibinfo {pages} {1} (\bibinfo {year} {2010})}\BibitemShut {NoStop}%
\bibitem [{\citenamefont {Schuld}\ and\ \citenamefont
  {Petruccione}(2018)}]{schuld2018supervised}%
  \BibitemOpen
  \bibfield  {author} {\bibinfo {author} {\bibfnamefont {M.}~\bibnamefont
  {Schuld}}\ and\ \bibinfo {author} {\bibfnamefont {F.}~\bibnamefont
  {Petruccione}},\ }\href {https://books.google.es/books?id=1zpsDwAAQBAJ}
  {\emph {\bibinfo {title} {Supervised Learning with Quantum Computers}}},\
  Quantum Science and Technology\ (\bibinfo  {publisher} {Springer
  International Publishing},\ \bibinfo {year} {2018})\BibitemShut {NoStop}%
\bibitem [{\citenamefont {P{\'e}rez-Salinas}\ \emph {et~al.}(2020)\citenamefont
  {P{\'e}rez-Salinas}, \citenamefont {Cervera-Lierta}, \citenamefont
  {Gil-Fuster},\ and\ \citenamefont {Latorre}}]{perez2020data}%
  \BibitemOpen
  \bibfield  {author} {\bibinfo {author} {\bibfnamefont {A.}~\bibnamefont
  {P{\'e}rez-Salinas}}, \bibinfo {author} {\bibfnamefont {A.}~\bibnamefont
  {Cervera-Lierta}}, \bibinfo {author} {\bibfnamefont {E.}~\bibnamefont
  {Gil-Fuster}}, \ and\ \bibinfo {author} {\bibfnamefont {J.~I.}\ \bibnamefont
  {Latorre}},\ }\href@noop {} {\bibfield  {journal} {\bibinfo  {journal}
  {Quantum}\ }\textbf {\bibinfo {volume} {4}},\ \bibinfo {pages} {226}
  (\bibinfo {year} {2020})},\ \Eprint {http://arxiv.org/abs/1907.02085}
  {arXiv:1907.02085} \BibitemShut {NoStop}%
\bibitem [{\citenamefont {Schuld}\ \emph {et~al.}(2021)\citenamefont {Schuld},
  \citenamefont {Sweke},\ and\ \citenamefont {Meyer}}]{schuld2021effect}%
  \BibitemOpen
  \bibfield  {author} {\bibinfo {author} {\bibfnamefont {M.}~\bibnamefont
  {Schuld}}, \bibinfo {author} {\bibfnamefont {R.}~\bibnamefont {Sweke}}, \
  and\ \bibinfo {author} {\bibfnamefont {J.~J.}\ \bibnamefont {Meyer}},\
  }\href@noop {} {\bibfield  {journal} {\bibinfo  {journal} {Physical Review
  A}\ }\textbf {\bibinfo {volume} {103}},\ \bibinfo {pages} {032430} (\bibinfo
  {year} {2021})},\ \Eprint {http://arxiv.org/abs/2008.08605}
  {arXiv:2008.08605} \BibitemShut {NoStop}%
\bibitem [{\citenamefont {Bergholm}\ \emph {et~al.}(2018)\citenamefont
  {Bergholm}, \citenamefont {Izaac}, \citenamefont {Schuld}, \citenamefont
  {Gogolin}, \citenamefont {Alam}, \citenamefont {Ahmed}, \citenamefont
  {Arrazola}, \citenamefont {Blank}, \citenamefont {Delgado}, \citenamefont
  {Jahangiri} \emph {et~al.}}]{bergholm2018pennylane}%
  \BibitemOpen
  \bibfield  {author} {\bibinfo {author} {\bibfnamefont {V.}~\bibnamefont
  {Bergholm}}, \bibinfo {author} {\bibfnamefont {J.}~\bibnamefont {Izaac}},
  \bibinfo {author} {\bibfnamefont {M.}~\bibnamefont {Schuld}}, \bibinfo
  {author} {\bibfnamefont {C.}~\bibnamefont {Gogolin}}, \bibinfo {author}
  {\bibfnamefont {M.~S.}\ \bibnamefont {Alam}}, \bibinfo {author}
  {\bibfnamefont {S.}~\bibnamefont {Ahmed}}, \bibinfo {author} {\bibfnamefont
  {J.~M.}\ \bibnamefont {Arrazola}}, \bibinfo {author} {\bibfnamefont
  {C.}~\bibnamefont {Blank}}, \bibinfo {author} {\bibfnamefont
  {A.}~\bibnamefont {Delgado}}, \bibinfo {author} {\bibfnamefont
  {S.}~\bibnamefont {Jahangiri}},  \emph {et~al.},\ }\href@noop {} {\bibfield
  {journal} {\bibinfo  {journal} {arXiv:1811.04968}\ } (\bibinfo {year}
  {2018})},\ \Eprint {http://arxiv.org/abs/rXiv:1811.04968}
  {arXiv:rXiv:1811.04968} \BibitemShut {NoStop}%
\bibitem [{\citenamefont {Bradbury}\ \emph {et~al.}(2018)\citenamefont
  {Bradbury}, \citenamefont {Frostig}, \citenamefont {Hawkins}, \citenamefont
  {Johnson}, \citenamefont {Leary}, \citenamefont {Maclaurin}, \citenamefont
  {Necula}, \citenamefont {Paszke}, \citenamefont {Vander{P}las}, \citenamefont
  {Wanderman-{M}ilne},\ and\ \citenamefont {Zhang}}]{jax2018github}%
  \BibitemOpen
  \bibfield  {author} {\bibinfo {author} {\bibfnamefont {J.}~\bibnamefont
  {Bradbury}}, \bibinfo {author} {\bibfnamefont {R.}~\bibnamefont {Frostig}},
  \bibinfo {author} {\bibfnamefont {P.}~\bibnamefont {Hawkins}}, \bibinfo
  {author} {\bibfnamefont {M.~J.}\ \bibnamefont {Johnson}}, \bibinfo {author}
  {\bibfnamefont {C.}~\bibnamefont {Leary}}, \bibinfo {author} {\bibfnamefont
  {D.}~\bibnamefont {Maclaurin}}, \bibinfo {author} {\bibfnamefont
  {G.}~\bibnamefont {Necula}}, \bibinfo {author} {\bibfnamefont
  {A.}~\bibnamefont {Paszke}}, \bibinfo {author} {\bibfnamefont
  {J.}~\bibnamefont {Vander{P}las}}, \bibinfo {author} {\bibfnamefont
  {S.}~\bibnamefont {Wanderman-{M}ilne}}, \ and\ \bibinfo {author}
  {\bibfnamefont {Q.}~\bibnamefont {Zhang}},\ }\href
  {http://github.com/google/jax} {\enquote {\bibinfo {title} {{JAX}: composable
  transformations of {P}ython+{N}um{P}y programs},}\ } (\bibinfo {year}
  {2018})\BibitemShut {NoStop}%
\bibitem [{\citenamefont {Patti}\ \emph
  {et~al.}(2021{\natexlab{a}})\citenamefont {Patti}, \citenamefont {Kossaifi},
  \citenamefont {Yelin},\ and\ \citenamefont {Anandkumar}}]{patti2021tensorly}%
  \BibitemOpen
  \bibfield  {author} {\bibinfo {author} {\bibfnamefont {T.~L.}\ \bibnamefont
  {Patti}}, \bibinfo {author} {\bibfnamefont {J.}~\bibnamefont {Kossaifi}},
  \bibinfo {author} {\bibfnamefont {S.~F.}\ \bibnamefont {Yelin}}, \ and\
  \bibinfo {author} {\bibfnamefont {A.}~\bibnamefont {Anandkumar}},\
  }\href@noop {} {\bibfield  {journal} {\bibinfo  {journal} {arXiv:2112.10239}\
  } (\bibinfo {year} {2021}{\natexlab{a}})},\ \Eprint
  {http://arxiv.org/abs/2112.10239} {arXiv:2112.10239} \BibitemShut {NoStop}%
\bibitem [{\citenamefont {Harris}\ \emph {et~al.}(2020)\citenamefont {Harris},
  \citenamefont {Millman}, \citenamefont {van~der Walt}, \citenamefont
  {Gommers}, \citenamefont {Virtanen}, \citenamefont {Cournapeau},
  \citenamefont {Wieser}, \citenamefont {Taylor}, \citenamefont {Berg},
  \citenamefont {Smith}, \citenamefont {Kern}, \citenamefont {Picus},
  \citenamefont {Hoyer}, \citenamefont {van Kerkwijk}, \citenamefont {Brett},
  \citenamefont {Haldane}, \citenamefont {del R{\'{i}}o}, \citenamefont
  {Wiebe}, \citenamefont {Peterson}, \citenamefont {G{\'{e}}rard-Marchant},
  \citenamefont {Sheppard}, \citenamefont {Reddy}, \citenamefont {Weckesser},
  \citenamefont {Abbasi}, \citenamefont {Gohlke},\ and\ \citenamefont
  {Oliphant}}]{harris2020array}%
  \BibitemOpen
  \bibfield  {author} {\bibinfo {author} {\bibfnamefont {C.~R.}\ \bibnamefont
  {Harris}}, \bibinfo {author} {\bibfnamefont {K.~J.}\ \bibnamefont {Millman}},
  \bibinfo {author} {\bibfnamefont {S.~J.}\ \bibnamefont {van~der Walt}},
  \bibinfo {author} {\bibfnamefont {R.}~\bibnamefont {Gommers}}, \bibinfo
  {author} {\bibfnamefont {P.}~\bibnamefont {Virtanen}}, \bibinfo {author}
  {\bibfnamefont {D.}~\bibnamefont {Cournapeau}}, \bibinfo {author}
  {\bibfnamefont {E.}~\bibnamefont {Wieser}}, \bibinfo {author} {\bibfnamefont
  {J.}~\bibnamefont {Taylor}}, \bibinfo {author} {\bibfnamefont
  {S.}~\bibnamefont {Berg}}, \bibinfo {author} {\bibfnamefont {N.~J.}\
  \bibnamefont {Smith}}, \bibinfo {author} {\bibfnamefont {R.}~\bibnamefont
  {Kern}}, \bibinfo {author} {\bibfnamefont {M.}~\bibnamefont {Picus}},
  \bibinfo {author} {\bibfnamefont {S.}~\bibnamefont {Hoyer}}, \bibinfo
  {author} {\bibfnamefont {M.~H.}\ \bibnamefont {van Kerkwijk}}, \bibinfo
  {author} {\bibfnamefont {M.}~\bibnamefont {Brett}}, \bibinfo {author}
  {\bibfnamefont {A.}~\bibnamefont {Haldane}}, \bibinfo {author} {\bibfnamefont
  {J.~F.}\ \bibnamefont {del R{\'{i}}o}}, \bibinfo {author} {\bibfnamefont
  {M.}~\bibnamefont {Wiebe}}, \bibinfo {author} {\bibfnamefont
  {P.}~\bibnamefont {Peterson}}, \bibinfo {author} {\bibfnamefont
  {P.}~\bibnamefont {G{\'{e}}rard-Marchant}}, \bibinfo {author} {\bibfnamefont
  {K.}~\bibnamefont {Sheppard}}, \bibinfo {author} {\bibfnamefont
  {T.}~\bibnamefont {Reddy}}, \bibinfo {author} {\bibfnamefont
  {W.}~\bibnamefont {Weckesser}}, \bibinfo {author} {\bibfnamefont
  {H.}~\bibnamefont {Abbasi}}, \bibinfo {author} {\bibfnamefont
  {C.}~\bibnamefont {Gohlke}}, \ and\ \bibinfo {author} {\bibfnamefont {T.~E.}\
  \bibnamefont {Oliphant}},\ }\href {\doibase 10.1038/s41586-020-2649-2}
  {\bibfield  {journal} {\bibinfo  {journal} {Nature}\ }\textbf {\bibinfo
  {volume} {585}},\ \bibinfo {pages} {357} (\bibinfo {year}
  {2020})}\BibitemShut {NoStop}%
\bibitem [{\citenamefont {Huembeli}\ \emph {et~al.}(2022)\citenamefont
  {Huembeli}, \citenamefont {Chaudhary},\ and\ \citenamefont
  {MacCormack}}]{huembeli_patrick_2022}%
  \BibitemOpen
  \bibfield  {author} {\bibinfo {author} {\bibfnamefont {P.}~\bibnamefont
  {Huembeli}}, \bibinfo {author} {\bibfnamefont {S.}~\bibnamefont {Chaudhary}},
  \ and\ \bibinfo {author} {\bibfnamefont {I.}~\bibnamefont {MacCormack}},\
  }\href {https://gitlab.com/menten_ai_public/discrete-qgan-paper-supplement}
  {\enquote {\bibinfo {title} {{GitLab: Towards a scalable discrete quantum
  generative adversarial neural network}},}\ } (\bibinfo {year}
  {2022})\BibitemShut {NoStop}%
\bibitem [{\citenamefont {Kingma}\ and\ \citenamefont
  {Ba}(2014)}]{kingma2014adam}%
  \BibitemOpen
  \bibfield  {author} {\bibinfo {author} {\bibfnamefont {D.~P.}\ \bibnamefont
  {Kingma}}\ and\ \bibinfo {author} {\bibfnamefont {J.}~\bibnamefont {Ba}},\
  }\href@noop {} {\bibfield  {journal} {\bibinfo  {journal} {arXiv:1412.6980}\
  } (\bibinfo {year} {2014})},\ \Eprint {http://arxiv.org/abs/1412.6980}
  {arXiv:1412.6980} \BibitemShut {NoStop}%
\bibitem [{\citenamefont {Gili}\ \emph
  {et~al.}(2022{\natexlab{b}})\citenamefont {Gili}, \citenamefont
  {Hibat-Allah}, \citenamefont {Mauri}, \citenamefont {Ballance},\ and\
  \citenamefont {Perdomo-Ortiz}}]{gili2022quantum}%
  \BibitemOpen
  \bibfield  {author} {\bibinfo {author} {\bibfnamefont {K.}~\bibnamefont
  {Gili}}, \bibinfo {author} {\bibfnamefont {M.}~\bibnamefont {Hibat-Allah}},
  \bibinfo {author} {\bibfnamefont {M.}~\bibnamefont {Mauri}}, \bibinfo
  {author} {\bibfnamefont {C.}~\bibnamefont {Ballance}}, \ and\ \bibinfo
  {author} {\bibfnamefont {A.}~\bibnamefont {Perdomo-Ortiz}},\ }\href@noop {}
  {\bibfield  {journal} {\bibinfo  {journal} {arXiv:2207.13645}\ } (\bibinfo
  {year} {2022}{\natexlab{b}})},\ \Eprint {http://arxiv.org/abs/2207.13645}
  {arXiv:2207.13645} \BibitemShut {NoStop}%
\bibitem [{\citenamefont {McClean}\ \emph {et~al.}(2018)\citenamefont
  {McClean}, \citenamefont {Boixo}, \citenamefont {Smelyanskiy}, \citenamefont
  {Babbush},\ and\ \citenamefont {Neven}}]{mcclean2018barren}%
  \BibitemOpen
  \bibfield  {author} {\bibinfo {author} {\bibfnamefont {J.~R.}\ \bibnamefont
  {McClean}}, \bibinfo {author} {\bibfnamefont {S.}~\bibnamefont {Boixo}},
  \bibinfo {author} {\bibfnamefont {V.~N.}\ \bibnamefont {Smelyanskiy}},
  \bibinfo {author} {\bibfnamefont {R.}~\bibnamefont {Babbush}}, \ and\
  \bibinfo {author} {\bibfnamefont {H.}~\bibnamefont {Neven}},\ }\href@noop {}
  {\bibfield  {journal} {\bibinfo  {journal} {Nature communications}\ }\textbf
  {\bibinfo {volume} {9}},\ \bibinfo {pages} {1} (\bibinfo {year}
  {2018})}\BibitemShut {NoStop}%
\bibitem [{\citenamefont {Patti}\ \emph
  {et~al.}(2021{\natexlab{b}})\citenamefont {Patti}, \citenamefont {Najafi},
  \citenamefont {Gao},\ and\ \citenamefont {Yelin}}]{patti2021entanglement}%
  \BibitemOpen
  \bibfield  {author} {\bibinfo {author} {\bibfnamefont {T.~L.}\ \bibnamefont
  {Patti}}, \bibinfo {author} {\bibfnamefont {K.}~\bibnamefont {Najafi}},
  \bibinfo {author} {\bibfnamefont {X.}~\bibnamefont {Gao}}, \ and\ \bibinfo
  {author} {\bibfnamefont {S.~F.}\ \bibnamefont {Yelin}},\ }\href@noop {}
  {\bibfield  {journal} {\bibinfo  {journal} {Physical Review Research}\
  }\textbf {\bibinfo {volume} {3}},\ \bibinfo {pages} {033090} (\bibinfo {year}
  {2021}{\natexlab{b}})},\ \Eprint {http://arxiv.org/abs/2012.12658}
  {arXiv:2012.12658} \BibitemShut {NoStop}%
\bibitem [{\citenamefont {Cerezo}\ \emph {et~al.}(2021)\citenamefont {Cerezo},
  \citenamefont {Sone}, \citenamefont {Volkoff}, \citenamefont {Cincio},\ and\
  \citenamefont {Coles}}]{cerezo2021cost}%
  \BibitemOpen
  \bibfield  {author} {\bibinfo {author} {\bibfnamefont {M.}~\bibnamefont
  {Cerezo}}, \bibinfo {author} {\bibfnamefont {A.}~\bibnamefont {Sone}},
  \bibinfo {author} {\bibfnamefont {T.}~\bibnamefont {Volkoff}}, \bibinfo
  {author} {\bibfnamefont {L.}~\bibnamefont {Cincio}}, \ and\ \bibinfo {author}
  {\bibfnamefont {P.~J.}\ \bibnamefont {Coles}},\ }\href@noop {} {\bibfield
  {journal} {\bibinfo  {journal} {Nature communications}\ }\textbf {\bibinfo
  {volume} {12}},\ \bibinfo {pages} {1} (\bibinfo {year} {2021})},\ \Eprint
  {http://arxiv.org/abs/2001.00550} {arXiv:2001.00550} \BibitemShut {NoStop}%
\bibitem [{\citenamefont {{Cong}}\ \emph {et~al.}(2019)\citenamefont {{Cong}},
  \citenamefont {{Choi}},\ and\ \citenamefont {{Lukin}}}]{2019NatPh..15.1273C}%
  \BibitemOpen
  \bibfield  {author} {\bibinfo {author} {\bibfnamefont {I.}~\bibnamefont
  {{Cong}}}, \bibinfo {author} {\bibfnamefont {S.}~\bibnamefont {{Choi}}}, \
  and\ \bibinfo {author} {\bibfnamefont {M.~D.}\ \bibnamefont {{Lukin}}},\
  }\href {\doibase 10.1038/s41567-019-0648-8} {\bibfield  {journal} {\bibinfo
  {journal} {Nature Physics}\ }\textbf {\bibinfo {volume} {15}},\ \bibinfo
  {pages} {1273} (\bibinfo {year} {2019})},\ \Eprint
  {http://arxiv.org/abs/1810.03787} {arXiv:1810.03787 [quant-ph]} \BibitemShut
  {NoStop}%
\bibitem [{\citenamefont {Pesah}\ \emph {et~al.}(2021)\citenamefont {Pesah},
  \citenamefont {Cerezo}, \citenamefont {Wang}, \citenamefont {Volkoff},
  \citenamefont {Sornborger},\ and\ \citenamefont {Coles}}]{pesah2021absence}%
  \BibitemOpen
  \bibfield  {author} {\bibinfo {author} {\bibfnamefont {A.}~\bibnamefont
  {Pesah}}, \bibinfo {author} {\bibfnamefont {M.}~\bibnamefont {Cerezo}},
  \bibinfo {author} {\bibfnamefont {S.}~\bibnamefont {Wang}}, \bibinfo {author}
  {\bibfnamefont {T.}~\bibnamefont {Volkoff}}, \bibinfo {author} {\bibfnamefont
  {A.~T.}\ \bibnamefont {Sornborger}}, \ and\ \bibinfo {author} {\bibfnamefont
  {P.~J.}\ \bibnamefont {Coles}},\ }\href@noop {} {\bibfield  {journal}
  {\bibinfo  {journal} {Physical Review X}\ }\textbf {\bibinfo {volume} {11}},\
  \bibinfo {pages} {041011} (\bibinfo {year} {2021})},\ \Eprint
  {http://arxiv.org/abs/2011.02966} {arXiv:2011.02966} \BibitemShut {NoStop}%
\bibitem [{\citenamefont {Sharma}\ \emph {et~al.}(2022)\citenamefont {Sharma},
  \citenamefont {Cerezo}, \citenamefont {Cincio},\ and\ \citenamefont
  {Coles}}]{sharma2022trainability}%
  \BibitemOpen
  \bibfield  {author} {\bibinfo {author} {\bibfnamefont {K.}~\bibnamefont
  {Sharma}}, \bibinfo {author} {\bibfnamefont {M.}~\bibnamefont {Cerezo}},
  \bibinfo {author} {\bibfnamefont {L.}~\bibnamefont {Cincio}}, \ and\ \bibinfo
  {author} {\bibfnamefont {P.~J.}\ \bibnamefont {Coles}},\ }\href@noop {}
  {\bibfield  {journal} {\bibinfo  {journal} {Physical Review Letters}\
  }\textbf {\bibinfo {volume} {128}},\ \bibinfo {pages} {180505} (\bibinfo
  {year} {2022})},\ \Eprint {http://arxiv.org/abs/2005.12458}
  {arXiv:2005.12458} \BibitemShut {NoStop}%
\bibitem [{\citenamefont {Shende}\ \emph {et~al.}(2004)\citenamefont {Shende},
  \citenamefont {Markov},\ and\ \citenamefont {Bullock}}]{shende2004minimal}%
  \BibitemOpen
  \bibfield  {author} {\bibinfo {author} {\bibfnamefont {V.~V.}\ \bibnamefont
  {Shende}}, \bibinfo {author} {\bibfnamefont {I.~L.}\ \bibnamefont {Markov}},
  \ and\ \bibinfo {author} {\bibfnamefont {S.~S.}\ \bibnamefont {Bullock}},\
  }\href@noop {} {\bibfield  {journal} {\bibinfo  {journal} {Physical Review
  A}\ }\textbf {\bibinfo {volume} {69}},\ \bibinfo {pages} {062321} (\bibinfo
  {year} {2004})},\ \Eprint {http://arxiv.org/abs/quant-ph/0308033}
  {arXiv:quant-ph/0308033} \BibitemShut {NoStop}%
\bibitem [{\citenamefont {Rakyta}\ and\ \citenamefont
  {Zimbor{\'a}s}(2022)}]{rakyta2022approaching}%
  \BibitemOpen
  \bibfield  {author} {\bibinfo {author} {\bibfnamefont {P.}~\bibnamefont
  {Rakyta}}\ and\ \bibinfo {author} {\bibfnamefont {Z.}~\bibnamefont
  {Zimbor{\'a}s}},\ }\href@noop {} {\bibfield  {journal} {\bibinfo  {journal}
  {Quantum}\ }\textbf {\bibinfo {volume} {6}},\ \bibinfo {pages} {710}
  (\bibinfo {year} {2022})},\ \Eprint {http://arxiv.org/abs/2109.06770}
  {arXiv:2109.06770} \BibitemShut {NoStop}%
\bibitem [{\citenamefont {Mirza}\ and\ \citenamefont
  {Osindero}(2014)}]{mirza2014conditional}%
  \BibitemOpen
  \bibfield  {author} {\bibinfo {author} {\bibfnamefont {M.}~\bibnamefont
  {Mirza}}\ and\ \bibinfo {author} {\bibfnamefont {S.}~\bibnamefont
  {Osindero}},\ }\href@noop {} {\bibfield  {journal} {\bibinfo  {journal}
  {arXiv:1411.1784}\ } (\bibinfo {year} {2014})},\ \Eprint
  {http://arxiv.org/abs/1411.1784} {arXiv:1411.1784} \BibitemShut {NoStop}%
\end{thebibliography}%
\appendix
\section{Discriminator with balanced data}
\label{Appendix:Discrimnator}
\begin{figure}[]%
    \centering
    \subfloat{{\includegraphics[width=0.44\linewidth]{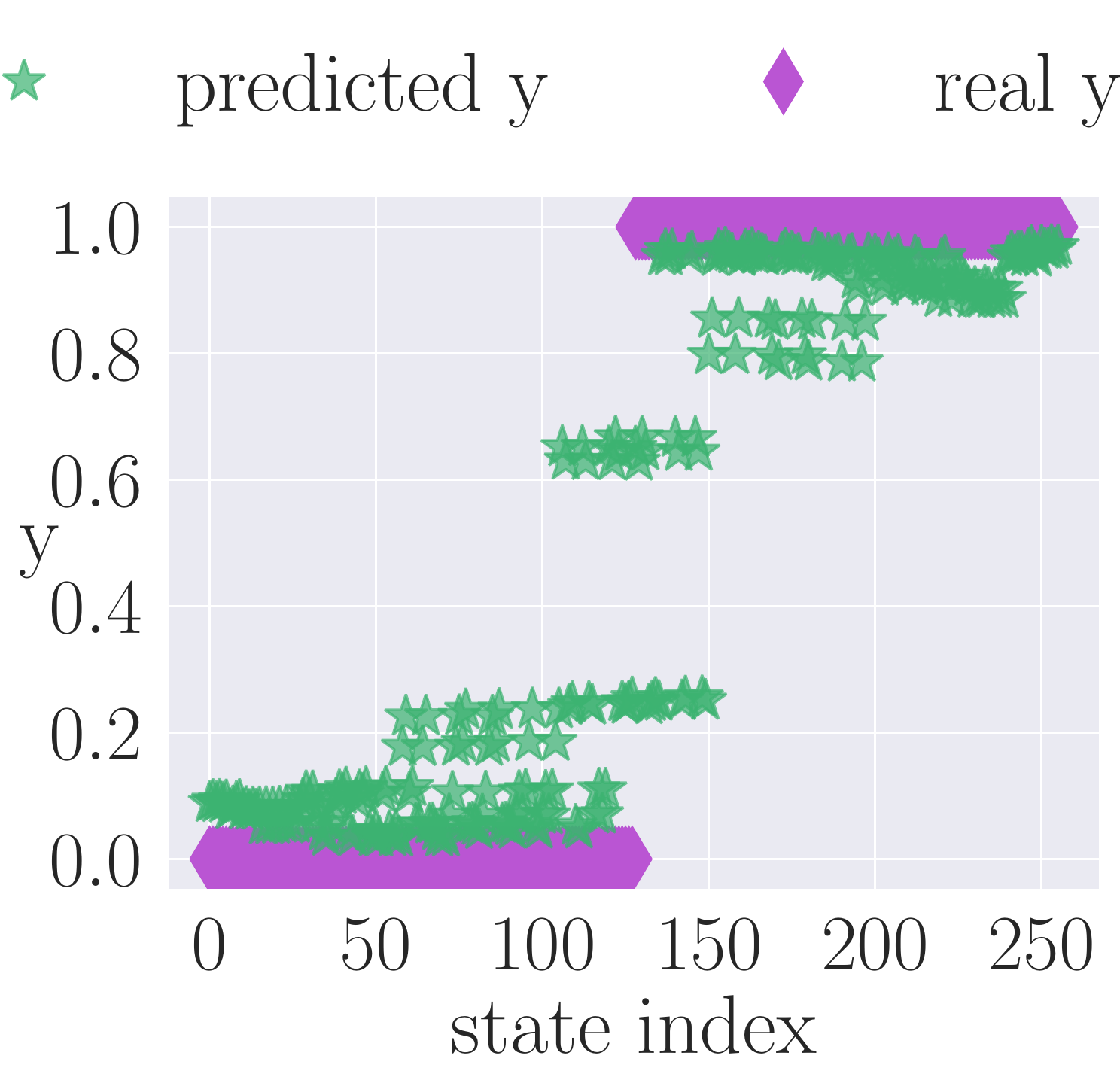} }}%
    \qquad
    \subfloat{{\includegraphics[width=0.44\linewidth]{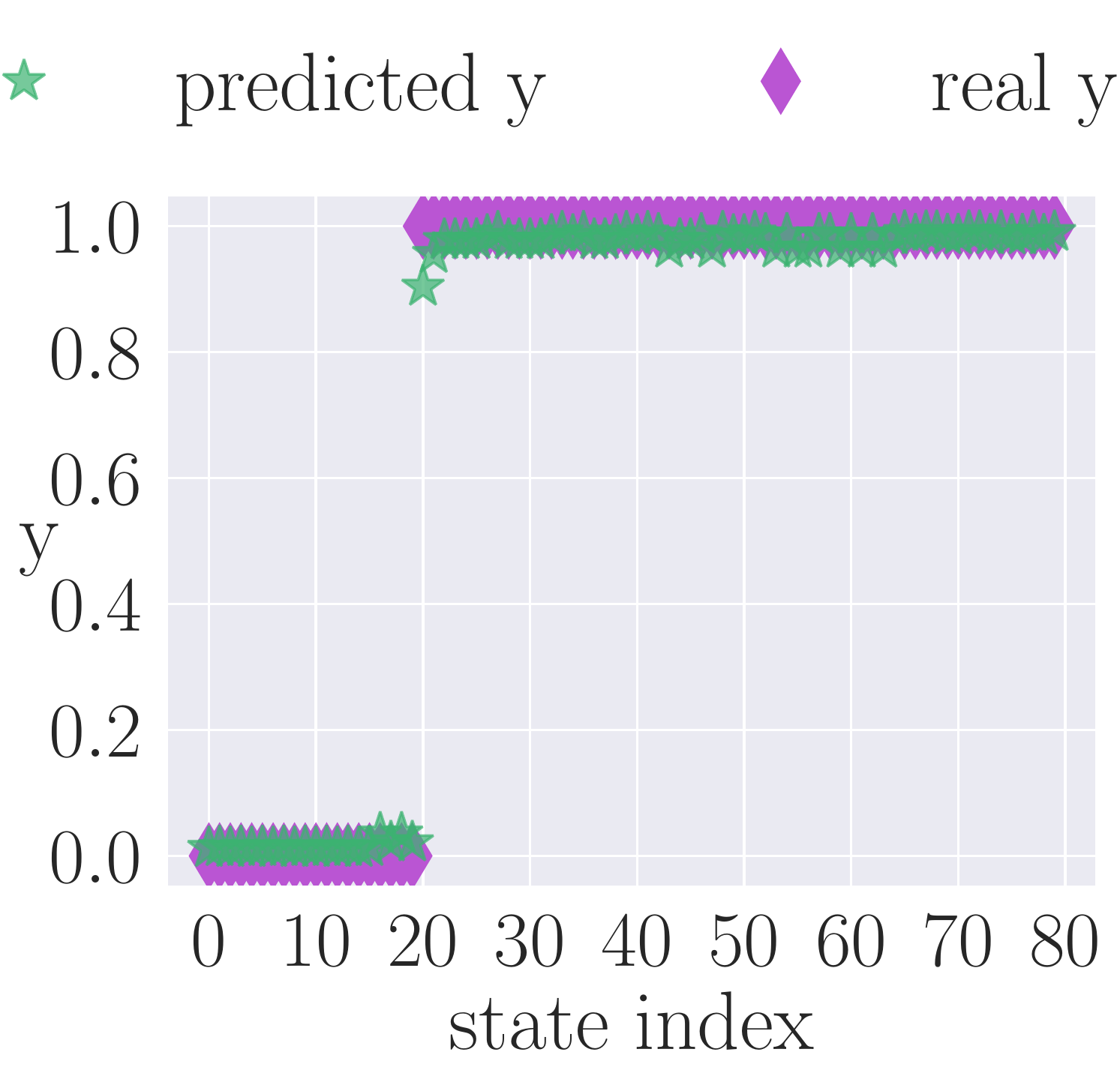} }}%
    \caption{\textbf{Discriminator balanced and reduced data:} Discriminator predictions vs. real labels for the Ising data on $N=8$ spins for a balanced dataset that contains all $2^N = 256$ states (left) and an imbalanced dataset that only consists of $80$ states with 20 low energy and 60 high energy states (right).}%
    \label{fig:Discriminator_Ising}%
\end{figure}

Fig.~\ref{fig:Discriminator_Ising} shows the discriminator's performance on the Ising dataset with $N=8$ spins and no auxiliary qubits. In the top panel, we compare the predicted and the real labels for a balanced dataset, where the whole state space of $2^N$ states is split into 2 equally sized sets of low energy ($y=0$) and high energy ($y=1$) states. One can see that the discriminator struggles to correctly classify some of the intermediate energy states but shows high accuracy in the low and high energy region.
Fig.~\ref{fig:Discriminator_Ising} (bottom) shows a scenario where $N=8$, but we only train the discriminator on a dataset of $20$ low energy states and it needs to distinguish them from $60$ random high energy states. To train a QGAN the discriminator will have to distinguish the training dataset, i.e. the low energy states, from some randomly sampled high energy states. If we assume that during the training of the QGAN we do not sample the whole space of $2^N$ states but only $60$ the discriminator shows high prediction accuracy even without auxiliary qubits.
\section{Toy model without noise: Single run}
\label{sec:no_noise_single}
\begin{figure}
    \centering
    \includegraphics[width=0.9\linewidth]{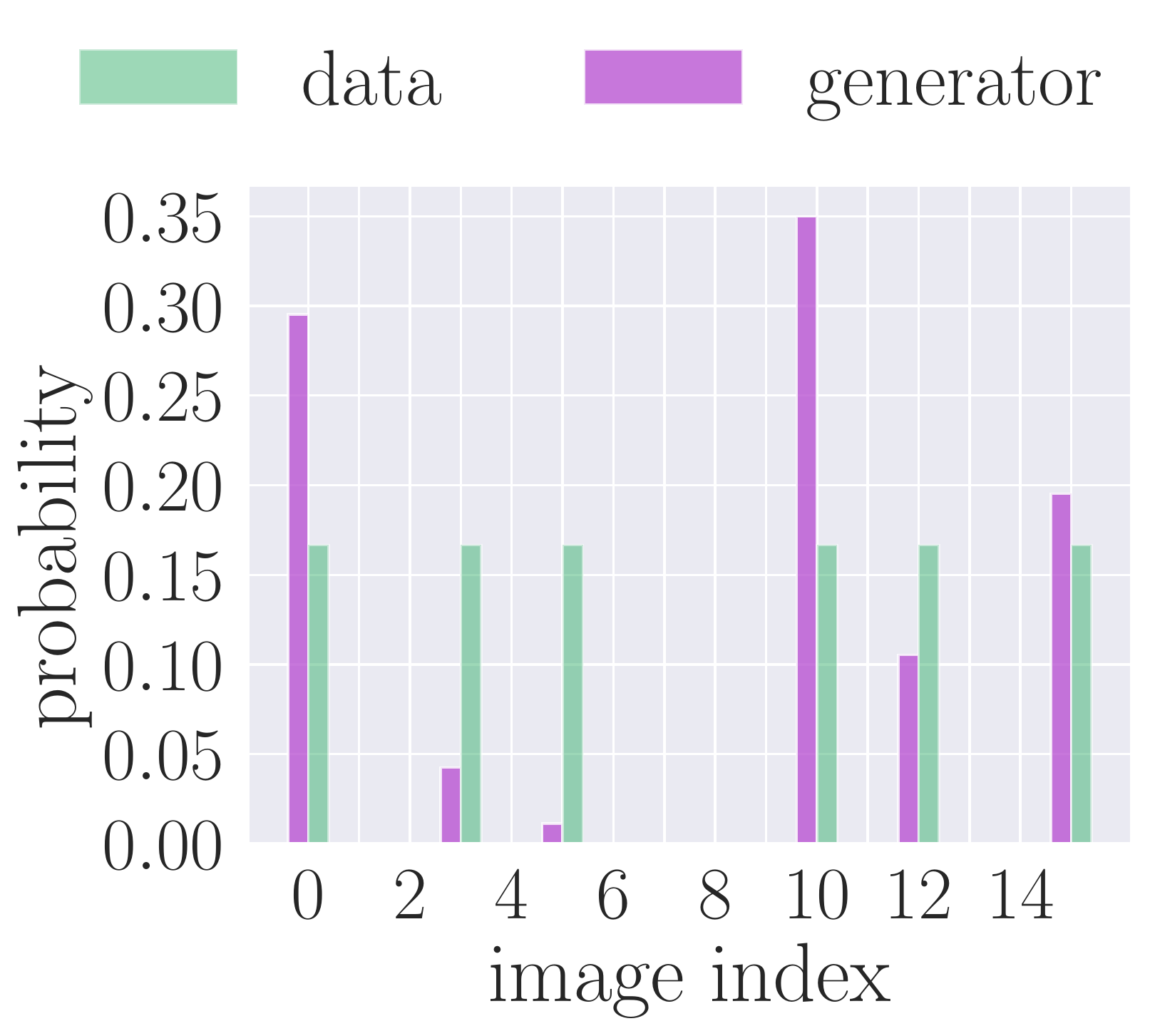}
    \caption{\textbf{Quantum generator toy model without noise (without averaging):} Output distribution of the generator quantum state $\vert \psi_G \rangle = \sum c_i \vert x_i \rangle$ without classical noise input, trained together with the discriminator circuit $U_D(\theta_D)$. As shown in Fig~\ref{fig:Toy_Model_no_noise}, the loss gets stuck very fast and the generator does not learn the training distribution. This figure shows the output distribution after a single training run without averaging. The training gets stuck and the images of the B\&S dataset with index 3 and 5 are barely learned, while others have too high frequency. This phenomenon is known as mode collapse.}
    \label{fig:no_noise_single}
\end{figure}

In Fig.~\ref{fig:Toy_Model_no_noise} the output distribution of the generator are averaged over $20$ training runs for different random initializations. The average distribution itself seems to be close to the data distribution, but the variance of the probabilities is far bigger than for the toy model with noise for example. We here show in Fig.~\ref{fig:no_noise_single} the output distribution of a single training run. One can see that the images 3 and 5 have very low probabilities and others have too high probabilities. This is a so-called data collapse. Since it is dependent on the initialization which of the data collapses, the average distribution over many training runs seems to be close to the actual data distribution.

\section{Linear noise model}
\label{sec:Appendix_linear_noise}
\begin{figure}[ht]%
    \centering
    \subfloat{{\includegraphics[width=0.44\linewidth]{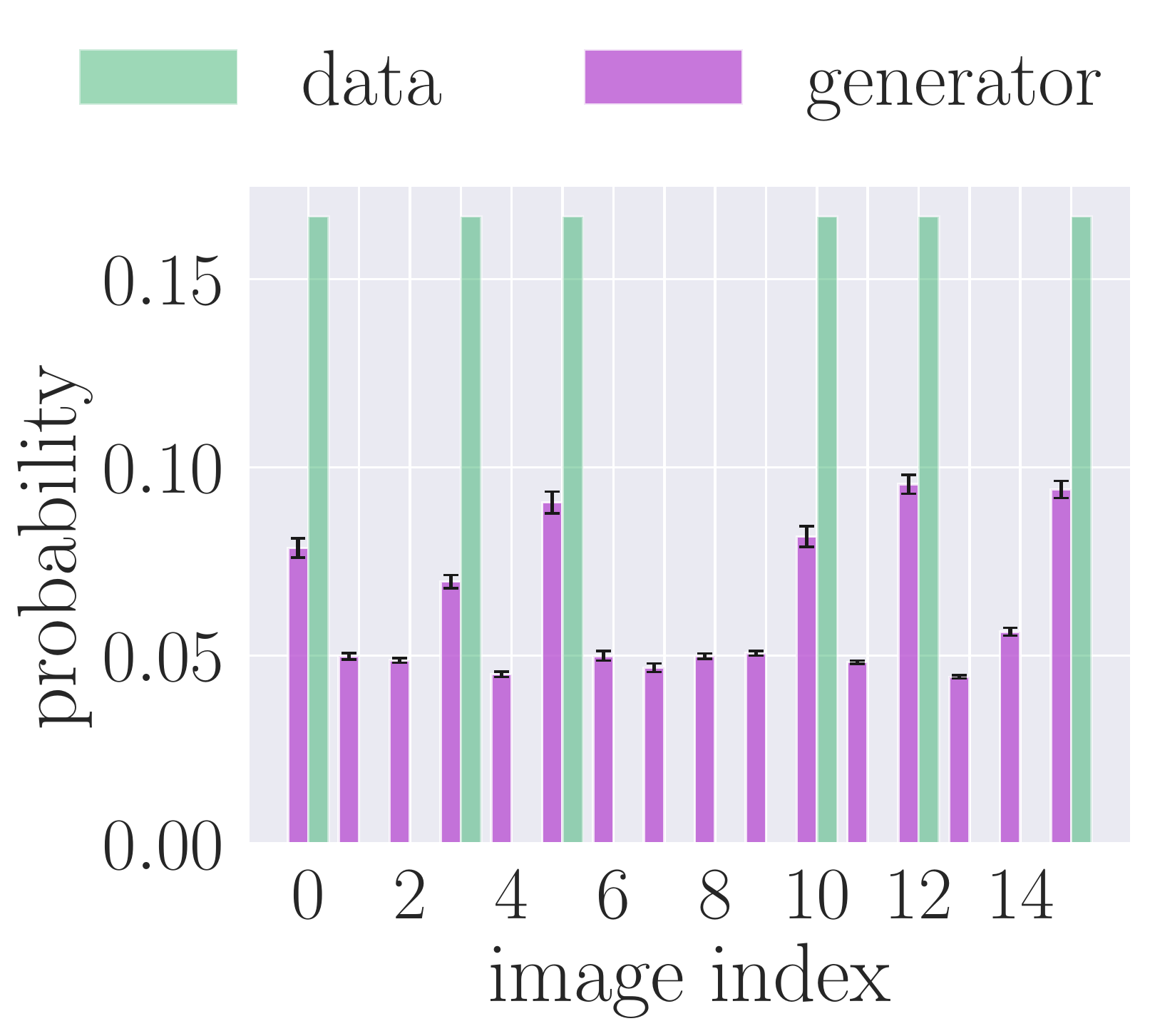} }}%
    \qquad
    \subfloat{{\includegraphics[width=0.44\linewidth]{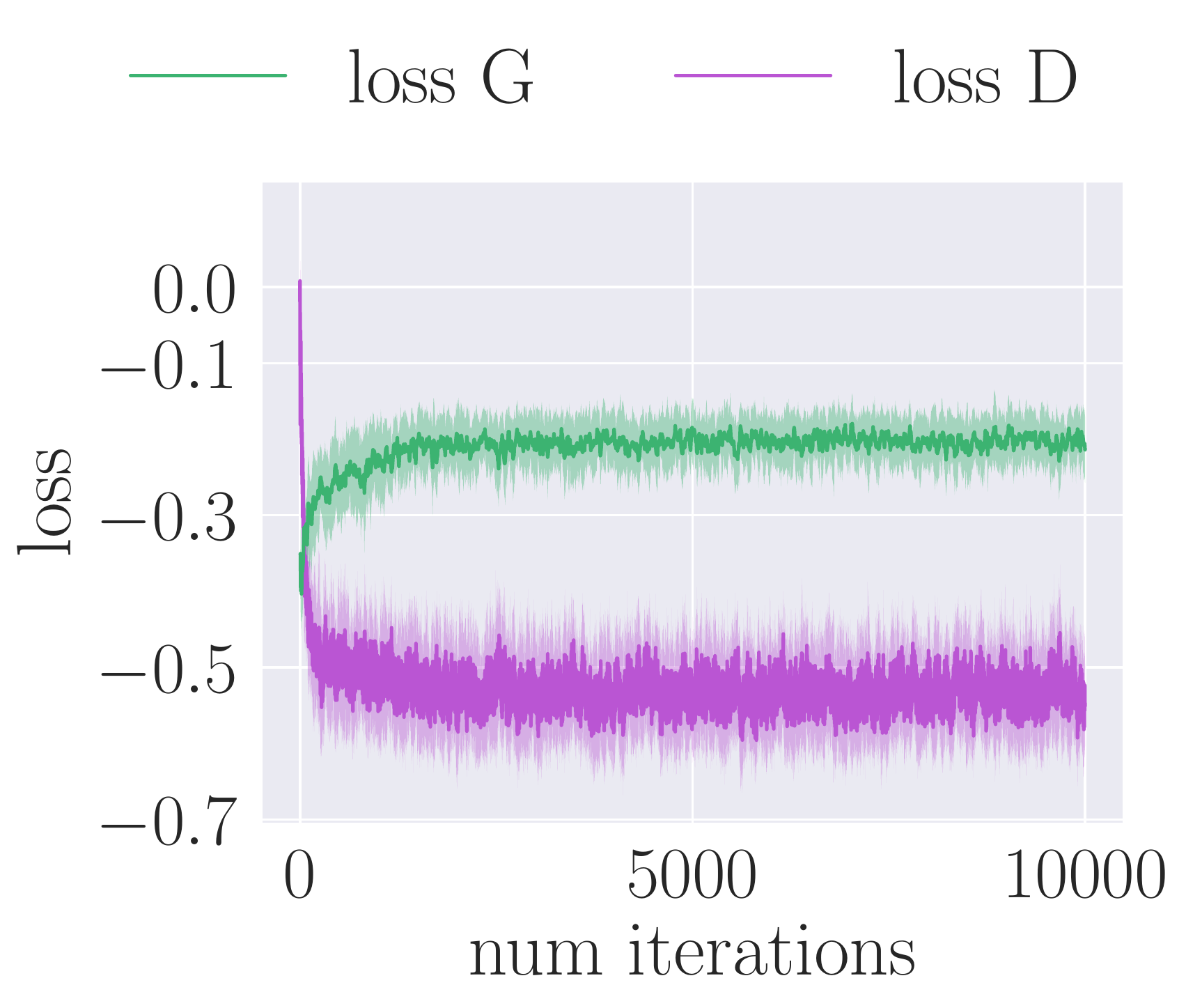} }}%
    \cprotect\caption{\textbf{Quantum generator with linear noise input:} Output distribution (left) and loss during training (right) of the generator quantum circuit with a noise state as an input $U_G(\theta_G)\ket{z}$ trained together with the discriminator circuit $U_D(\theta_D)$. The training does not converge as reliably as with the reuploading circuit but the output distribution of the generator comes close to the training data. Remarkably, the loss functions are not converging to the expected Nash equilibrium. \verb|loss_G| should converge to $-0.5$, not to $0$ and \verb|loss_D| to $0$. Both figures are averaged over 20 different random initializations. The error bars in the histogram show the variance of the probabilities. The shaded region in the right panel shows the standard deviation.}%
    \label{fig:QNN_linear_noise}%
\end{figure}

Fig.~\ref{fig:QNN_linear_noise} shows the training of a fully quantum GAN without data reuploading but with a noise model where the generator state reads $U_G(\theta_G) U(z) \ket{0}^{\otimes N}$. Even though the output distribution of the generator comes close to the one with data reuploading, it seems like the QGAN does not converge to the correct distribution and consequently samples the images that are not in the training set too high and the rest too low. Furthermore, the loss curves do not converge to the correct Wasserstein GAN losses. This behaviour can be reproduced for most choices of noise and circuit depth. See also our Gitlab repository~\cite{huembeli_patrick_2022}.
\end{document}